%% file: t2k-p0d-pi0-rate-v2.tex
\documentclass[aps,prd,showpacs,superscriptaddress,groupedaddress,showpacs,showkeys]{revtex4-1}  
\pdfoutput=1

\usepackage{lineno}
\usepackage{graphicx}
\usepackage{amssymb,amsmath}
\usepackage{morefloats}
\usepackage{url}
\usepackage{subfigure}

\newcommand{\tableIndentOne}[1]{\qquad{}}
\newcommand{\tableIndentTwo}[1]{\qquad{}\qquad{}}

\newcommand{\pizero}{\ensuremath{\pi^0}}
\newcommand{\ncpi}{NC1\pizero{}}
\newcommand{\sys}{\ensuremath{\textrm{\,(sys.)}}}
\newcommand{\stat}{\ensuremath{\textrm{\,(stat.)}}}
\newcommand{\flux}{\ensuremath{\textrm{\,(flux)}}}
\newcommand{\MeVcc}{\ensuremath{\textrm{MeV} / \textrm{c}^{2}}}

\newcommand{\NCPiNeutrinoPeak}{\ensuremath{0.6}~GeV}

\newcommand{\NCPiNeutrinoEnergy}{\ensuremath{1.3}~GeV}

\newcommand{\NCPiNeutrinoFluxSys}{\ensuremath{12 \%}}

\newcommand{\MCOnWaterCrossSection}{\ensuremath{7.63 \times 10^{-39} ~ \textrm{cm}^2}}
\newcommand{\MCOnWaterNucleonCrossSection}{\ensuremath{4.24 \times 10^{-40} ~\textrm{cm}^2~\textrm{nucleon}^{-1}}} 
 
\newcommand{\MCWaterOutAMUCrossSection}{\ensuremath{4.20 \times 10^{-40} ~\textrm{cm}^2~\textrm{amu}^{-1}}} 

\newcommand{\DataWaterIPOT}{\ensuremath{2.96\times 10^{19}}}
\newcommand{\DataWaterIIPOT}{\ensuremath{6.96\times 10^{19}}}
\newcommand{\DataAirIIPOT}{\ensuremath{3.59\times 10^{19}}}

\newcommand{\DataAirIIIPOT}{\ensuremath{1.35\times 10^{20}}}
\newcommand{\DataWaterIVPOT}{\ensuremath{1.65\times 10^{20}}}
\newcommand{\DataAirIVPOT}{\ensuremath{1.78\times 10^{20}}}

\newcommand{\DataWaterPOT}{\ensuremath{2.64\times 10^{20}}}
\newcommand{\DataAirPOT}{\ensuremath{3.49\times 10^{20}}}

\newcommand{\MCWaterPOT}{\ensuremath{7.18\times 10^{21}}}
\newcommand{\MCAirPOT}{\ensuremath{4.01\times 10^{21}}}

\newcommand{\DataWaterInSelected}{\ensuremath{775}}
\newcommand{\MCWaterInSelected}{\ensuremath{893}}
\newcommand{\DataWaterOutSelected}{\ensuremath{555}}
\newcommand{\MCWaterOutSelected}{\ensuremath{629}}

\newcommand{\DataWaterInMuDk}{\ensuremath{227}}
\newcommand{\MCWaterInMuDk}{\ensuremath{331}}
\newcommand{\DataWaterOutMuDk}{\ensuremath{123}}
\newcommand{\MCWaterOutMuDk}{\ensuremath{210}}

\newcommand{\DataWaterInSigValue}{\ensuremath{342}}
\newcommand{\DataWaterInSigStat}{\ensuremath{33}}
\newcommand{\DataWaterInSigSys}{\ensuremath{88}}
\newcommand{\DataWaterInSigFit}{\ensuremath{\DataWaterInSigValue{} 
    \pm{} \DataWaterInSigStat{}}}
\newcommand{\DataWaterInSig}{\ensuremath{\DataWaterInSigFit{} \stat{}
    \pm{}\DataWaterInSigSys{}\sys}}
\newcommand{\MCWaterInSig}{\ensuremath{433}}

\newcommand{\DataWaterOutSigValue}{\ensuremath{246}}
\newcommand{\DataWaterOutSigStat}{\ensuremath{26}}
\newcommand{\DataWaterOutSigSys}{\ensuremath{61}}
\newcommand{\DataWaterOutSigFit}{\ensuremath{\DataWaterOutSigValue{} 
    \pm{} \DataWaterOutSigStat{}}}
\newcommand{\DataWaterOutSig}{\ensuremath{\DataWaterOutSigFit{} \stat{}
    \pm{} \DataWaterOutSigSys{} \sys{}}}
\newcommand{\MCWaterOutSig}{\ensuremath{290}}

\newcommand{\DataOnWaterSigValue}{\ensuremath{106}}
\newcommand{\DataOnWaterSigStat}{\ensuremath{41}}
\newcommand{\DataOnWaterSigSys}{\ensuremath{69}}
\newcommand{\DataOnWaterSig}{\ensuremath{\DataOnWaterSigValue{}
    \pm{} \DataOnWaterSigStat{} \stat{}
    \pm{} \DataOnWaterSigSys \sys{}}}
\newcommand{\MCOnWaterSig}{\ensuremath{157}}

\newcommand{\DataMCWaterInRatioValue}{\ensuremath{0.79}}
\newcommand{\DataMCWaterInRatioStat}{\ensuremath{0.08}}
\newcommand{\DataMCWaterInRatioSys}{\ensuremath{0.20}}
\newcommand{\DataMCWaterInRatio}{\ensuremath{\DataMCWaterInRatioValue{} 
    \pm{} \DataMCWaterInRatioStat{} \stat{}
    \pm{} \DataMCWaterInRatioSys{} \sys{}}}
\newcommand{\DataMCWaterOutRatioValue}{\ensuremath{0.85}}
\newcommand{\DataMCWaterOutRatioStat}{\ensuremath{0.09}}
\newcommand{\DataMCWaterOutRatioSys}{\ensuremath{0.21}}
\newcommand{\DataMCWaterOutRatio}{\ensuremath{\DataMCWaterOutRatioValue{} 
    \pm{} \DataMCWaterOutRatioStat{} \stat{}
    \pm{} \DataMCWaterOutRatioSys{} \sys{}}}
\newcommand{\DataMCOnWaterRatioValue}{\ensuremath{0.68}}
\newcommand{\DataMCOnWaterRatioStat}{\ensuremath{0.26}}
\newcommand{\DataMCOnWaterRatioSys}{\ensuremath{0.44}}
\newcommand{\DataMCOnWaterRatioFlux}{\ensuremath{0.12}}

\newcommand{\DataMCOnWaterFluxRatio}{\ensuremath{\DataMCOnWaterRatioValue{} 
    \pm{} \DataMCOnWaterRatioStat{} \stat{}
    \pm{} \DataMCOnWaterRatioSys{} \sys{}
    \pm{} \DataMCOnWaterRatioFlux{} \flux{}}}

\newcommand{\WaterInVtxXMean}{\ensuremath{-0.06}}
\newcommand{\WaterInVtxXSigma}{\ensuremath{5.5}}
\newcommand{\WaterInVtxYMean}{\ensuremath{0.06}}
\newcommand{\WaterInVtxYSigma}{\ensuremath{6.1}}
\newcommand{\WaterInVtxZMean}{\ensuremath{1.67}}
\newcommand{\WaterInVtxZSigma}{\ensuremath{8.6}}

\newcommand{\WaterOutVtxXMean}{\ensuremath{0.08}}
\newcommand{\WaterOutVtxXSigma}{\ensuremath{6.8}}
\newcommand{\WaterOutVtxYMean}{\ensuremath{0.20}}
\newcommand{\WaterOutVtxYSigma}{\ensuremath{8.0}}
\newcommand{\WaterOutVtxZMean}{\ensuremath{1.72}}
\newcommand{\WaterOutVtxZSigma}{\ensuremath{11.2}}

\newcommand{\WaterInMomMean}{\ensuremath{-3.2 \%}}
\newcommand{\WaterInMomSigma}{\ensuremath{18.7 \%}}
\newcommand{\WaterOutMomMean}{\ensuremath{-0.8 \%}}
\newcommand{\WaterOutMomSigma}{\ensuremath{21.1 \%}}

\newcommand{\GeometryVarSys}{\ensuremath{2.8 \%}}

\newcommand{\WaterInEnergyScaleSys}{\ensuremath{5.8 \%}}
\newcommand{\WaterOutEnergyScaleSys}{\ensuremath{0.9 \%}}

\newcommand{\WaterInPEPeakSys}{\ensuremath{0.6 \%}}
\newcommand{\WaterOutPEPeakSys}{\ensuremath{0.4 \%}}

\newcommand{\TimeVariationSys}{\ensuremath{1.8 \%}}

\newcommand{\WaterOutMassSys}{\ensuremath{0.6 \%}}
\newcommand{\WaterInMassSys}{\ensuremath{0.4 \%}}

\newcommand{\AlignmentSys}{\ensuremath{< 0.1 \%}}

\newcommand{\WaterInFiducialVolumeSys}{\ensuremath{1.5 \%}}
\newcommand{\WaterOutFiducialVolumeSys}{\ensuremath{2.0 \%}}

\newcommand{\WaterInFiducialShiftSys}{\ensuremath{1.1 \%}}
\newcommand{\WaterOutFiducialShiftSys}{\ensuremath{1.7 \%}}

\newcommand{\WaterInFluxSys}{\ensuremath{2.5 \%}}
\newcommand{\WaterOutFluxSys}{\ensuremath{3.3 \%}}

\newcommand{\WaterInTrackPIDSys}{\ensuremath{5.4 \%}}
\newcommand{\WaterOutTrackPIDSys}{\ensuremath{5.1 \%}}

\newcommand{\WaterInShowerSepSys}{\ensuremath{9.1 \%}}
\newcommand{\WaterOutShowerSepSys}{\ensuremath{11.6 \%}}

\newcommand{\WaterInShowerPIDSys}{\ensuremath{6.6 \%}}
\newcommand{\WaterOutShowerPIDSys}{\ensuremath{2.6 \%}}

\newcommand{\WaterInShowerChargeSys}{\ensuremath{6.6 \%}}
\newcommand{\WaterOutShowerChargeSys}{\ensuremath{3.0 \%}}

\newcommand{\WaterBpBConstraint}{\ensuremath{3.4 \%}}
\newcommand{\AirBpBConstraint}{\ensuremath{5.2 \%}}

\newcommand{\WaterSpSConstraint}{\ensuremath{\pm{} 1.6 \%}}
\newcommand{\AirSpSConstraint}{\ensuremath{\pm{} 2.0 \%}}

\newcommand{\WaterInGFactorStat}{\ensuremath{3.8 \%}}
\newcommand{\WaterOutGFactorStat}{\ensuremath{4.2 \%}}

\newcommand{\CorrelatedGFactorSys}{\ensuremath{19.8 \%}}

\newcommand{\WaterInTotalUncorrelatedSys}{\ensuremath{16.4 \%}}
\newcommand{\WaterOutTotalUncorrelatedSys}{\ensuremath{15.0 \%}}

\newcommand{\TotalCorrelatedSys}{\CorrelatedGFactorSys{}}

\newcommand{\WaterInTotalSystematic}{\ensuremath{25.7 \%}}
\newcommand{\WaterOutTotalSystematic}{\ensuremath{24.8 \%}}

\begin{document}

\title{Measurement of the single \pizero{} production rate in neutral current neutrino interactions on water}

\input{author_list.tex}

\begin{abstract}
The single \pizero{} production rate in neutral current neutrino interactions on water in a neutrino beam with a peak neutrino energy of \NCPiNeutrinoPeak{} has been measured using the P\O{}D, one of the subdetectors of the T2K near detector.
The production rate was measured for data taking periods when the P\O{}D contained water (\DataWaterPOT{}~protons-on-target) and also periods without water (\DataAirPOT{}~protons-on-target).
A measurement of the neutral current single \pizero{} production rate on water is made using appropriate subtraction of the production rate with water in from the rate with water out of the target region.
The subtraction analysis yields \DataOnWaterSig{} signal events, which is consistent with the prediction of \MCOnWaterSig{} events from the nominal simulation.
The measured to expected ratio is \DataMCOnWaterFluxRatio{}.
The nominal simulation uses a flux integrated cross section of \MCOnWaterCrossSection{} per nucleon with an average neutrino interaction energy of \NCPiNeutrinoEnergy{}.
\end{abstract}

\maketitle

\section{Introduction}

The Tokai to Kamioka (T2K) long-baseline neutrino experiment is designed to make precision measurements of the neutrino oscillation parameters $\theta_{23}$ and $\Delta m^2_{32}$ via $\nu_{\mu}$ disappearance and to search for the mixing angle $\theta_{13}$ via $\nu_e$ appearance in a $\nu_{\mu}$ beam.
An intense, almost pure beam of $\nu_{\mu}$ is produced by colliding 30 GeV protons with a graphite target at the J-PARC facility in Tokai-mura, Ibaraki\cite{Abe:2011ks}.
The resultant neutrino beam is directed $2.5^{\circ}$ away from the axis between the target and the far detector, resulting in a narrow band beam with peak energy near 0.6 GeV.
The direction, stability and flux of the beam are measured using a suite of near detectors (ND280) located 280 m downstream of the target.
At this distance, the neutrino beam is not expected to have been affected by oscillations.
The far detector, Super-Kamiokande, is located 295 km downstream of the target, a distance consistent with the oscillation maximum.
Super-Kamiokande uses water as both a detection medium and target to measure the amount of $\nu_e$ and $\nu_{\mu}$ present after oscillation  has occurred.
As neutral current \pizero{} events can cause an irreducible background to the $\nu_e$ appearance signal, it is important to provide a constraint using measurements of the production rate on water using the near detector.

This paper details the first measurement of neutral current single \pizero{} production (\ncpi{}) on water, using a neutrino beam with peak energy of ~\NCPiNeutrinoPeak{}\cite{Abe:2012av}.
The mean neutrino energy for the \ncpi{} interactions selected in this analysis is \NCPiNeutrinoEnergy{}.

Two processes dominate neutral current single \pizero{} production by neutrinos: resonant production and coherent scattering.
In resonant production, a neutrino interacts with a nucleon to produce a baryonic resonance, usually $\Delta(1232)$, which subsequently decays to a nucleon and a \pizero{}.
Coherent scattering occurs when a neutrino interacts with the entire nucleus, exchanging little energy and leaving the nucleus in its ground state.
The dominant decay mode for a \pizero{} is to two photons \cite{Agashe:2014kda} and if one decay photon is not detected, an \ncpi{} event can be indistinguishable from a charged current $\nu_e$ interaction, leading to an irreducible background in $\nu_{\mu} \to \nu_e$ oscillation measurements.
Whilst previous measurements performed using the T2K near detector have improved our knowledge of sub-GeV neutrino interactions, the rate of \ncpi{} production on water is still relatively unknown at the neutrino energies of the T2K beam.
Measurements of \ncpi{} production on a variety of targets and different neutrino energy distributions have been made at other experiments\cite{Pohl:1979ya,Nakayama:2004dp,AguilarArevalo:2009ww,Kurimoto:2009wq,Kurimoto:2010rc}.

In this analysis, the signal is defined by the final state particles, with an \ncpi{} interaction defined by a single \pizero{} particle exiting the nucleus along with any number of protons and neutrons but no charged leptons or other mesons.  
The rate of signal events on water is determined using event samples with a two photon signature from exposures with water in and out of the target region.
Using the presence of a muon decay tag, two photon candidate events are divided into signal-enriched and background-enriched samples.
The number of signal events, number of background events, energy scale, and shape of the background are then determined for each sample using a simultaneous maximum likelihood fit to the invariant mass distribution of the signal-enriched and background-enriched samples.
Finally, the number of interactions on water is determined using a weighted subtraction of the rate of signal events determined during the exposure with water in the target region and the exposure with the water removed.

The major sections of this paper are as follows.
Section~\ref{sec:detector} describes the T2K ND280 \pizero{} detector, as well as the simulation of the expected neutrino interactions and detector response.
Section~\ref{sec:eventselection} describes the event selection efficiencies and reconstruction resolutions for signal-enriched and background-enriched event samples and the selected event samples are described in Section~\ref{sec:selectedsample}.
The extraction of the number of signal events is described in Section~\ref{sec:signalextraction} followed by a discussion of the systematic uncertainty in Section~\ref{sec:systematics}.
Section~\ref{sec:eventrate} describes the calculation of the event rate on water and compares it with the expectation.

\section{\label{sec:detector} T2K ND280 P\O{}D Description and Simulation}
The T2K ND280 \pizero{} Detector (P\O{}D) is a scintillator-based tracking calorimeter optimized to measure \ncpi{} production in the momentum range that contributes backgrounds to $\nu_e$ appearance measurements~\cite{Assylbekov:2011sh}.
The P\O{}D is composed of layers of plastic scintillator alternating with water bags and brass or lead and is one of the first large scale detectors to use Multi-Pixel Photon Counters (MPPCs).
Relative to the neutrino beam, it sits upstream of a tracking detector made up of two fine grain scintillator modules placed between three time projection chambers.
Both the P\O{}D and tracking detector are in a 0.2~T magnetic field and surrounded by electromagnetic calorimeters and muon range detectors \cite{Abgrall:2010hi,Allan:2013ofa,Aoki:2012mf,Amaudruz:2012agx}.

The P\O{}D comprises 40 scintillator modules, each 38~mm thick, formed from two layers of scintillating bars with the long axis oriented either horizontally, or vertically, and instrumented using wavelength shifting fibers with an MPPC on one end and mirrored on the other~\cite{Assylbekov:2011sh}.
The triangular scintillating bars used to produce each of the two layers in each module have a height of 17~mm and a base of 32~mm and are interlocked to form a layer that is 17~mm thick.  
Two views are formed of an event, commonly labeled the X-Z, and the Y-Z view, where the $z$~axis is horizontal and points downstream, the $y$~axis points in the vertical direction, and the $x$~axis is perpendicular to the Y-Z plane.
A minimum ionizing particle will typically generate a charge in the MPPC equivalent to approximately $38~\mathrm{photoelectrons}/\mathrm{cm}$, or an average of about $30~\mathrm{photoelectrons}$ in a single bar.
The scintillator modules are arranged in three regions.
The most upstream and downstream regions are made of seven modules interleaved with 4.5~mm thick sheets of stainless steel-clad lead that function as 4.9 radiation length electromagnetic calorimeters to improve the containment of photons and electrons.
The central region serves as a target containing water.
It has 25 water target layers that are 28~mm thick interleaved with 26 scintillator modules and 1.3~mm brass sheets.
When water is in the detector, the target fiducial region contains approximately 1900~kg of water and 3570~kg of other materials.
Data collected with and without water in the P\O{}D are analyzed separately.

\begin{table}
  \centering
  \caption{\label{tab:datapot} Summary of T2K runs, including the
    configuration of the P\O{}D, and the number of protons on target
    (POT) used in this analysis.}
  \begin{ruledtabular}
    \begin{tabular}[c]{lcc}
      T2K Run & P\O{}D Configuration & POT \\
      \hline
      Run I   & Water-In  & \DataWaterIPOT  \\
      Run II  & Water-In  & \DataWaterIIPOT  \\
      Run II  & Water-Out & \DataAirIIPOT \\
      Run III & Water-Out & \DataAirIIIPOT \\
      Run IV  & Water-In  & \DataWaterIVPOT \\
      Run IV  & Water-Out & \DataAirIVPOT \\
      \hline
      Total   & Water-Out  & \DataAirPOT{} \\
      Total   & Water-In  & \DataWaterPOT{} \\
    \end{tabular}
  \end{ruledtabular}
\end{table}

This analysis utilizes data collected with a predominantly $\nu_\mu$ beam generated between January 2010 and May 2013, see \cite{Abe:2012av} for a detailed description.
The neutrinos are generated using a fast extracted $30~\mathrm{GeV}$ proton beam with a spill of 6--8 bunches that are separated by 582~ns.  
The proton beam strikes a graphite target producing pions and kaons which, after magnetically focusing the positive mesons, decay in flight to neutrinos.
The magnetic focusing can be altered to focus negative mesons.
The T2K runs, the configuration of the P\O{}D, and the corresponding protons on target (POT) are summarized in Table~\ref{tab:datapot}.

\label{sec:simulation} The simulated data set used in this analysis corresponds to \MCAirPOT{}~POT (water-out configuration), and \MCWaterPOT{}~POT (water-in configuration).
Neutrino interactions are simulated using the NEUT~\cite{Hayato:2009zz} event generator, version 5.1.4.2, with the interactions distributed within the full ND280 volume, as well as the surrounding hall.
Interactions on all nuclear targets present in ND280 are simulated.
Details of the neutrino interaction simulation process are described in~\cite{Hayato:2009zz}, \cite{PhysRevD.87.092003} and \cite{Abe:2014usb}.
The T2K run periods are simulated using the nominal detector and beam configurations, and then combined using the appropriate POT normalization to form the final expectation.
External, non-beam associated, backgrounds are not simulated, but are limited in the data sample by the duty cycle of the neutrino beam.
Particles produced in neutrino interactions are simulated using GEANT 4.9.4 \cite{Agostinelli:2002hh}.
The standard GEANT physics list for electromagnetic interactions is used in the simulation.

Neutral current single \pizero{} production in the T2K neutrino beam is dominated by resonant $\Delta(1232)$ production, which is simulated using the Rein-Sehgal \cite{Rein:1980wg} model for neutrino-induced resonant pion production.
The simulated \ncpi{} cross section on water integrated over the T2K neutrino beam flux is \MCOnWaterCrossSection{} or \MCOnWaterNucleonCrossSection{} while the \ncpi{} cross section for the fiducial region in the water-out configuration is \MCWaterOutAMUCrossSection{}~\cite{Abe:2012av,Hayato:2009zz}.  
There is an additional \NCPiNeutrinoFluxSys{} uncertainty in the neutrino flux integrated over the energy of neutrinos generating an \ncpi{} interaction\cite{Abe:2012av}.
This uncertainty is larger than presented in other T2K analyses because of the higher average neutrino energy, and, for this analysis, is unconstrained by other near detector measurements to allow direct comparison between the data and simulation.

\section{\label{sec:eventselection}Event Reconstruction and Selection}
Events are reconstructed in the P\O{}D using scintillation light signals that occur in time windows containing the neutrino bunch arrival.
A hit is constructed from the integrated charge during each time window, and the time relative to the start of the window at which the integrated light signal crosses a threshold equivalent to approximately 2.5~photoelectrons.
Activity in different time windows is independently reconstructed as separate events.
The reconstruction proceeds by selecting groups of hits consistent with a track-like signature that is classified as a light-ionizing track, such as a muon or charged pion, a heavy-ionizing track, such as a proton, or a non-track object such as a portion of an electromagnetic shower.
Hits from non-track objects, as well as any hits not gathered into a track-like object are then used to form groups that are consistent with shower-like particles such as photons or electrons coming from a single vertex.
In events without a track-like signature, the vertex is estimated by assuming that the particle signatures in the event emanate from a single point, with the particle directions going away from the vertex.
While events with track-like objects will generally be rejected in the later analysis, if a track-like object is found, then the vertex is fixed at the upstream end of the longest track.
After vertex reconstruction, all reconstructed non-track like objects, are  classified as either EM-like, or shower-like.
The shower-like objects primarily comprise interacting pions, interacting protons, or misidentified light-ionizing tracks.
The result of the reconstruction is a single vertex with an associated collection of objects corresponding to light-ionizing tracks, heavy-ionizing tracks, EM-like, and other shower-like objects.
A muon decay tag is associated with the reconstructed vertex when energy deposition consistent with a Michel electron is found.

A signal-enriched sample of exactly two photon candidates with invariant mass less than 500~\MeVcc{} is selected using eight selection criteria: event quality, vertex in the fiducial volume, energy containment in the P\O{}D, lack of a muon decay signature, fraction of energy in the two most energetic photon candidates, particle identification, reconstructed direction and object separation.  
In comparison to the signal, a distinguishing characteristic of the background is that it contains either a $\mu$, or a charged pion, both of which can generate a muon decay signature, so a separate background-enriched sample is selected by applying all criteria with the exception of the muon decay criterion which is reversed.

To be considered in this analysis, an event must occur during a neutrino beam spill and have a single reconstructed vertex as well as good data quality.  
The vertex must be in the fiducial volume defined as at least $25$ cm from the edge of the active volume and inside the water target region of the P\O{}D \cite{thesis:gilje}.
The containment criteria requires that all reconstructed objects are contained inside the P\O{}D by requiring that no reconstructed objects have hits in the last layer of the P\O{}D or in the outer two bars of any layer. 
This limits external background and improves the photon energy reconstruction.

The signature of interest is two reconstructed photons from the \pizero{} decay with no evidence of a muon-like object.
To ensure that selected events have two reconstructed photon candidates containing most of the recorded energy deposition, a ``charge-in-shower'' requirement is placed on the fraction of energy in the two most energetic EM-like objects.
The required fractions of 92\% (water-in), and 80\% (water-out) were chosen to optimize the statistical significance of the selected number of signal events using simulated samples.
Due to the planar nature of the P\O{}D, and the shape of the scintillator bars, the performance degrades for particles at an angle of more than approximately $75^{\circ}$ from the $z$~axis.
As such, the direction of the reconstructed total event momentum must be less than $60^\circ$ from the $z$~axis, limiting the phase space covered by this measurement.

Two well-separated decay photon candidates are required to limit the background from particles with overlapping energy deposits. 
The object separation in each projection is calculated by finding the distance between the two closest hits of the reconstructed objects.
Due to the planar nature of the P\O{}D, it is possible for two objects to overlap in one projection, but not in the other and separation is only required in one of the two projections.
The object separation is required to be greater than 9~cm (14~cm) in at least one projection for the water-in (water-out) configuration.

\begin{table}
  \centering
  \caption{\label{tab:effsummary} 
Efficiencies and the purity of the selection.  On-Water and Not-Water indicate the material that the neutrino interacted with.}
  \begin{ruledtabular}
    \begin{tabular}[c]{lcc} 
      & Efficiency, $\epsilon$ & Purity \\
      \hline
      Water-In& & \\
      \quad Total & $6.10 \%$ & $48.7 \%$\\
      \qquad On-Water & $6.20 \%$ & $56.2 \%$\\
      \qquad Not-Water & $6.04 \%$ & $45.3 \%$\\
      \hline
      Water-Out\hfill & & \\
      \quad Total & $4.79 \%$ & $46.1 \%$ \\
    \end{tabular}
  \end{ruledtabular}
\end{table}

\begin{figure}
  \centering
  \subfigure[Water-In configuration]{
    \includegraphics[width=0.45\textwidth]{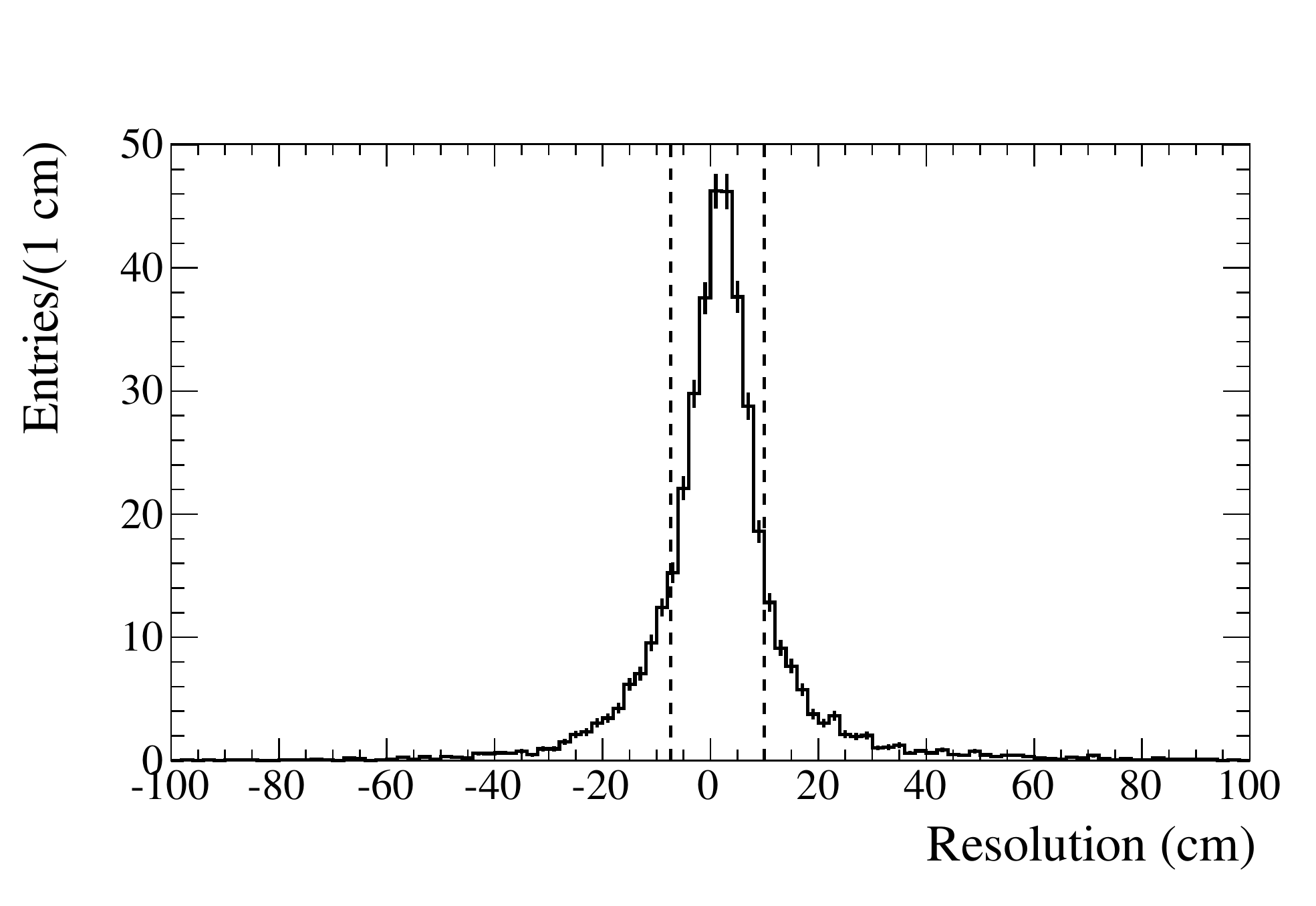}}
  \subfigure[Water-Out configuration]{
    \includegraphics[width=0.45\textwidth]{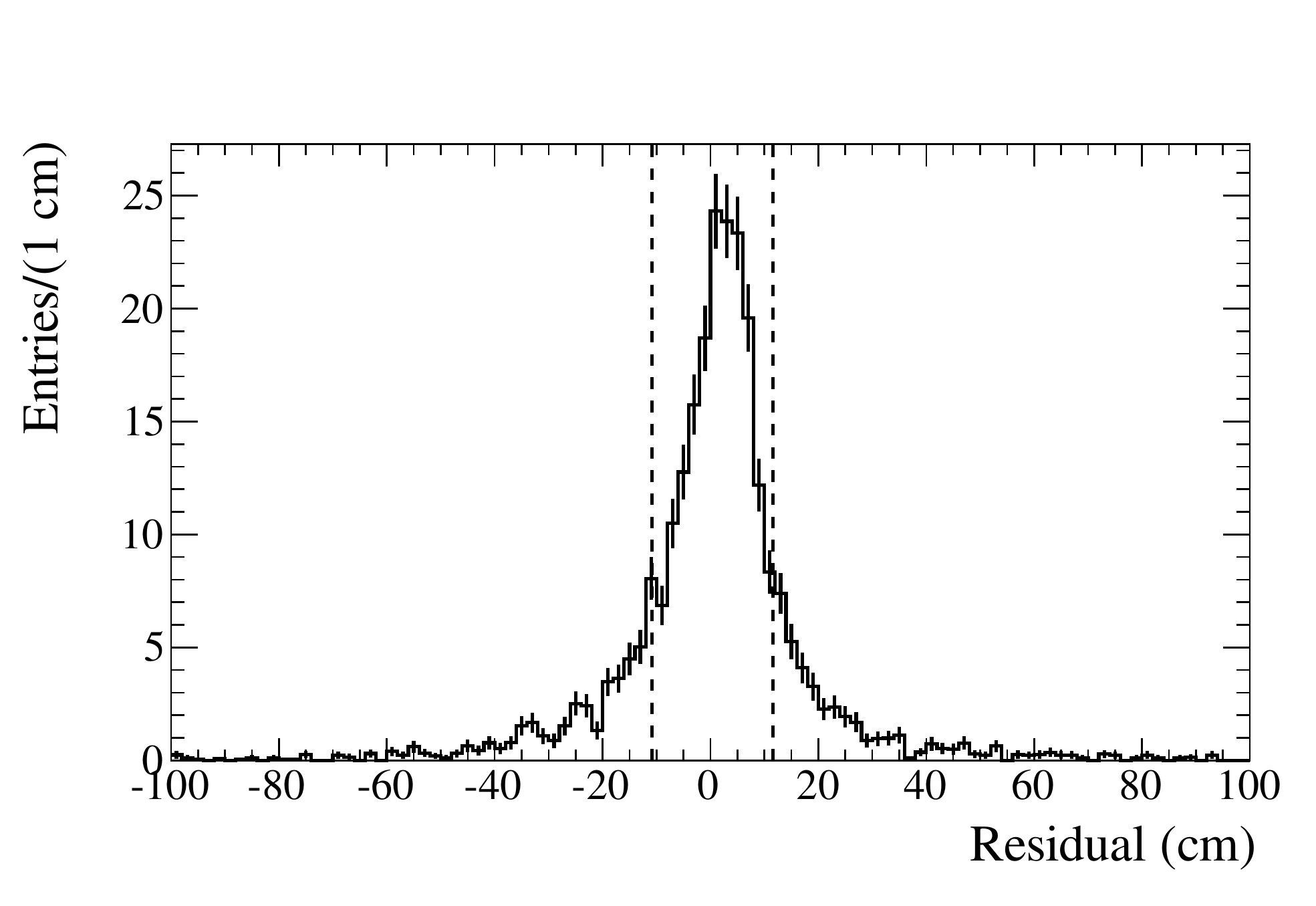}}
  \caption{\label{fig:vtxZRes}
The residual of the reconstructed vertex position relative to the true vertex position along the beam direction for selected \ncpi{} events.
The vertical lines correspond to the $16 \%$ and $84 \%$ quantiles.
}
\end{figure}

\begin{figure}
  \centering
  \subfigure[Water-In configuration]{
    \includegraphics[width=0.45\textwidth]{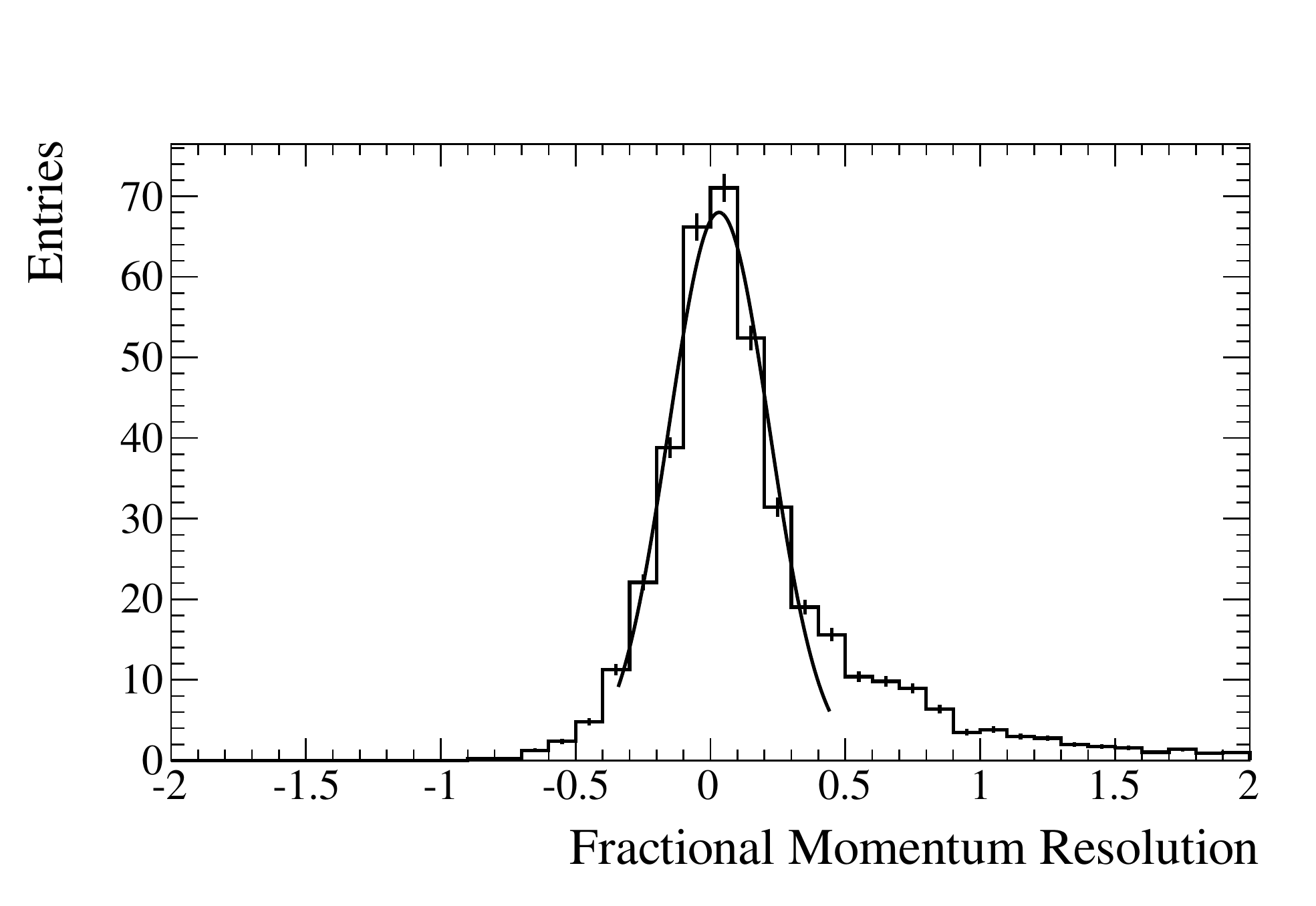}}
  \subfigure[Water-Out configuration]{
    \includegraphics[width=0.45\textwidth]{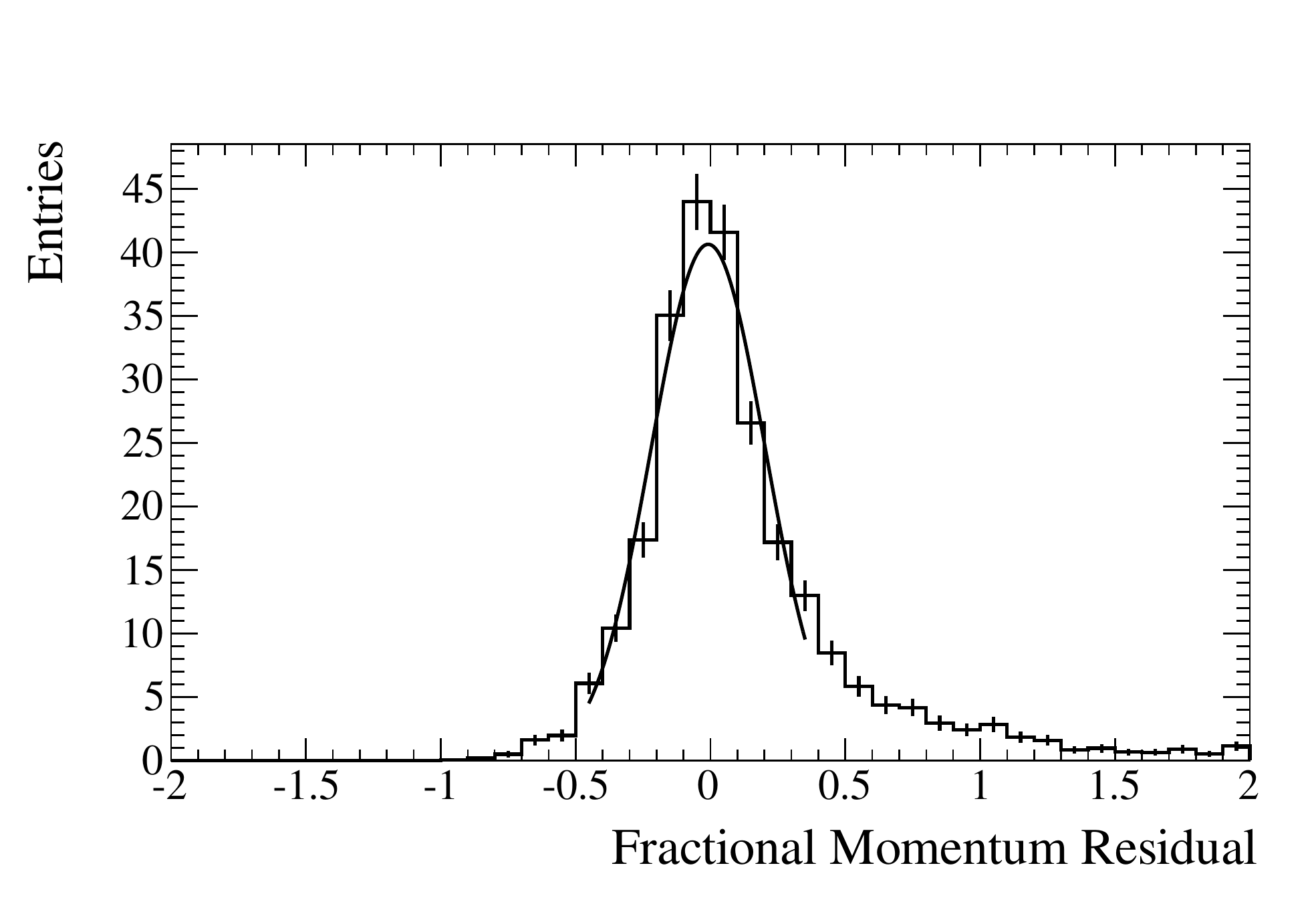}}
  \caption{\label{fig:momRes}
The distribution of the fractional difference between the reconstructed \ncpi{} momentum and the true momentum for selected \ncpi{} events.
A Gaussian distribution is fit to the central region.}
\end{figure}

Figure~\ref{fig:vtxZRes} shows the position of the reconstructed vertex relative to the true vertex position along the $z$~axis for the water-out configuration of the P\O{}D for simulated \ncpi{} events that have passed all selection criteria.
For the water-in (water-out) configuration, the biases are $\WaterInVtxXMean{}$~cm ($\WaterOutVtxXMean{}$~cm) along the $x$~axis, $\WaterInVtxYMean{}$~cm ($\WaterOutVtxYMean{}$~cm) along the $y$~axis, and $\WaterInVtxZMean{}$~cm ($\WaterOutVtxZMean{}$~cm) along the $z$~axis.
The expected vertex residual distribution is asymmetric due it's dependence on the reconstructed photon position and direction, and is characterized by half the distance between the $16 \%$ and $84 \%$ quantiles.
For the water-in (water-out) configuration, the resolutions are $\WaterInVtxXSigma{}$~cm ($\WaterOutVtxXSigma{}$~cm) along the $x$~axis, $\WaterInVtxYSigma{}$~cm ($\WaterOutVtxYSigma{}$~cm) along the $y$~axis,
$\WaterInVtxZSigma{}$~cm ($\WaterOutVtxZSigma{}$~cm) along the $z$~axis.

The momentum resolution of the \pizero{} is a combination of the energy and angular resolution for two reconstructed photons.
The total energy for the reconstructed photons is determined calorimetrically, and the fraction of the total energy carried by each reconstruct photons is calculated using the projections where the photon objects are geometrically distinct.
Figure~\ref{fig:momRes} shows the fractional momentum residual (the difference of the reconstructed and true momenta divided by the true momentum) for the water-out configuration of the P\O{}D for \ncpi{} events passing all selection criteria.
A Gaussian is fit to the central region to determine the shift and width of the momentum distribution.
The fractional momentum residual distribution has a mean of \WaterInMomMean{} with a width of \WaterInMomSigma{} for the P\O{}D water-in configuration.
For the water-out configuration, the mean is \WaterOutMomMean{} with a width of \WaterOutMomSigma{}.
The reconstructed opening angle distribution for simulated \ncpi{} events passing all selection criteria in both the water-in and water-out configurations has a mean of $-0.01$~rad from the nominal value and an RMS of $0.06$~rad.

\begin{figure}
  \centering
  \subfigure[Water-In configuration]{    \includegraphics[width=0.45\textwidth]{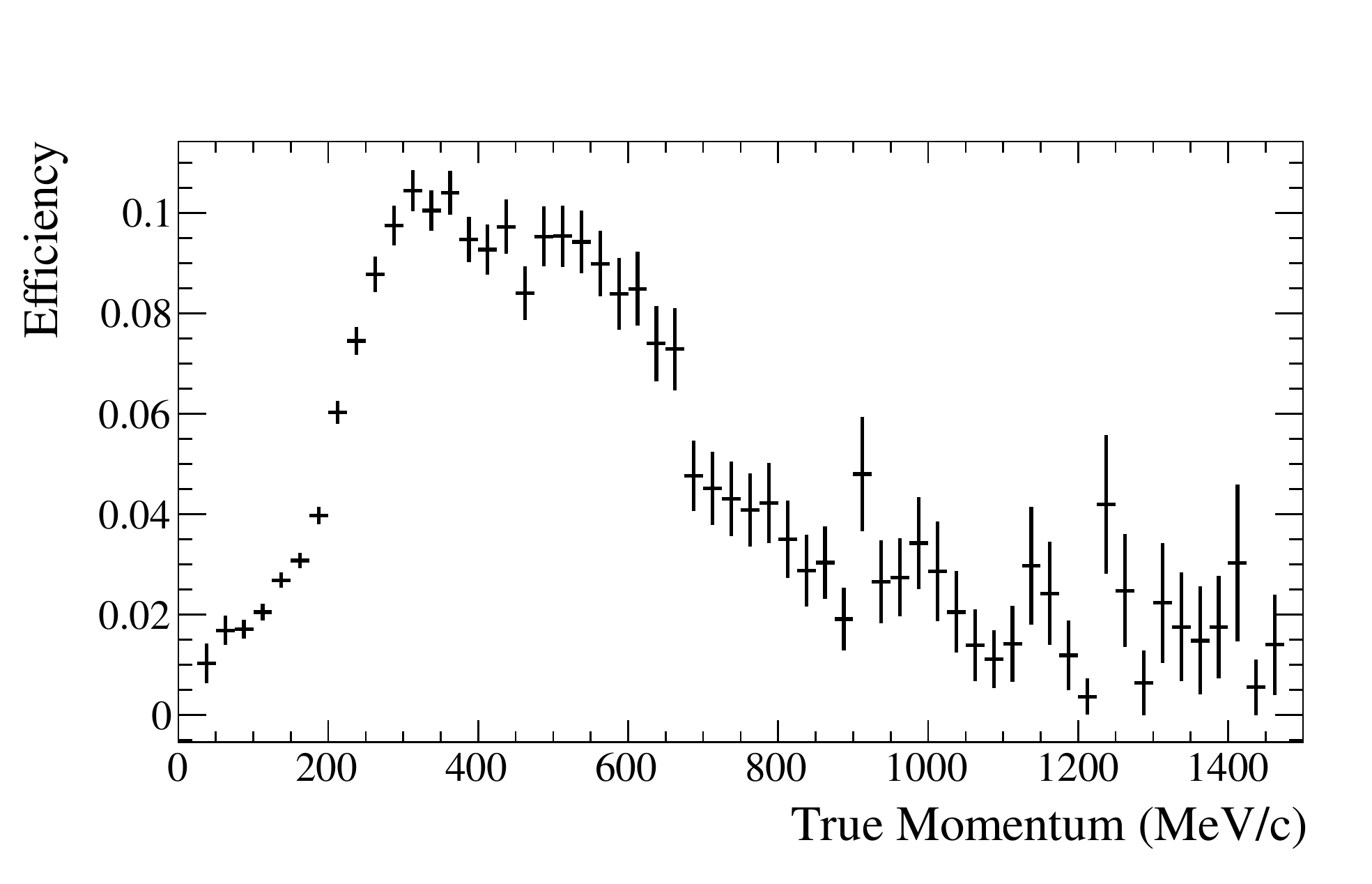}}
  \subfigure[Water-Out configuration]{      \includegraphics[width=0.45\textwidth]{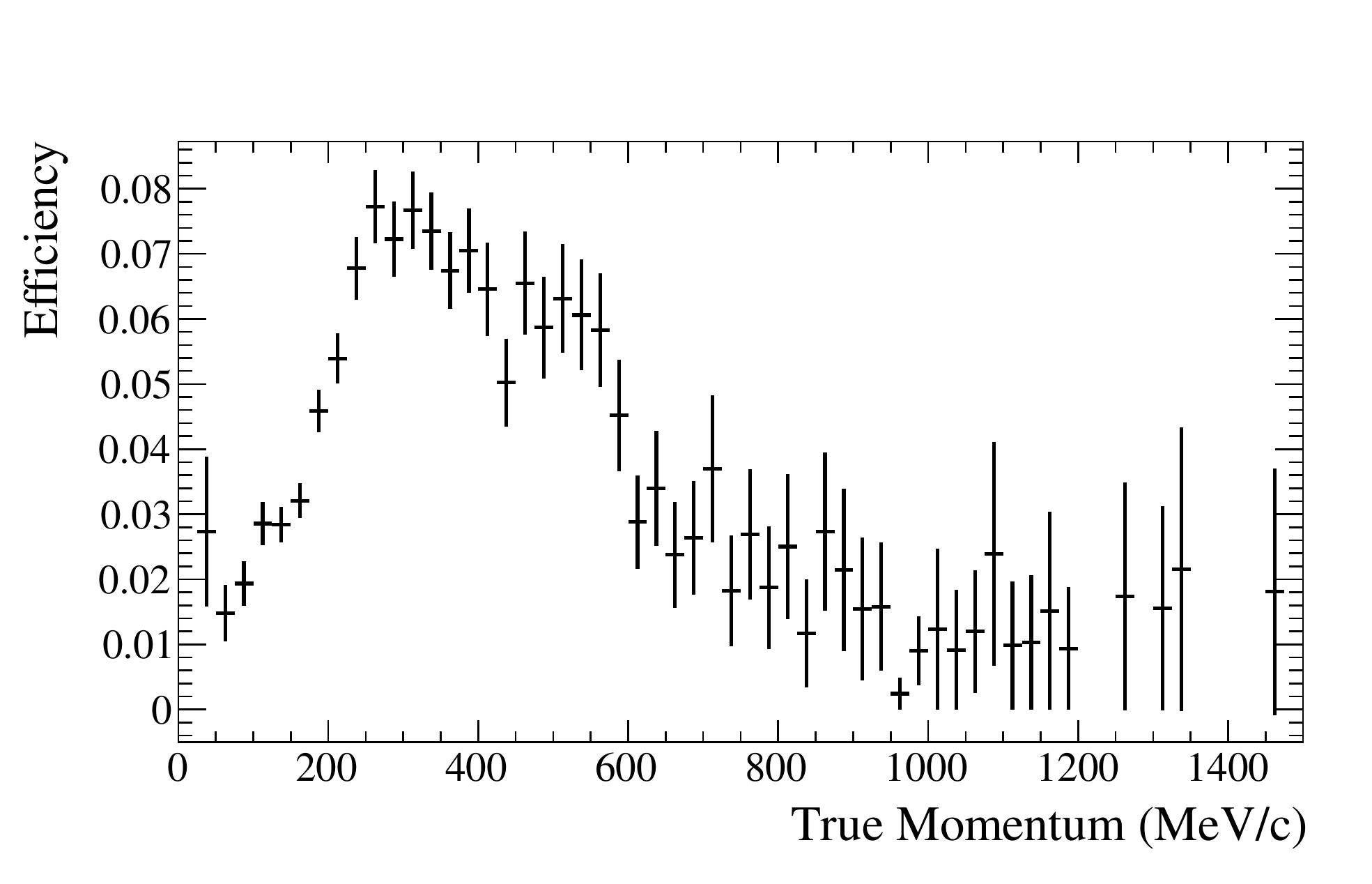}}
  \caption{\label{fig:effmom} The efficiency to select an \ncpi{} event as a
    function of the true momentum of the \pizero{}.}
\end{figure}

The reconstruction efficiency, $\epsilon$, for an \ncpi{} event and signal-enriched sample purity are summarized in Table~\ref{tab:effsummary}.
The efficiency is defined as the number of true \ncpi{} events reconstructed in the fiducial volume divided by the number of \ncpi{} interactions occurring within the same volume, while the purity is defined as the fraction of the selected events which result from a true \ncpi{} interaction.
The average efficiency is $6.10 \%$ for the water-in configuration, and $4.79 \%$ for the water-out configuration.
There is a small location dependence in the efficiency for the water-in configuration, so the efficiency is tabulated separately for interactions which occur in the water target, and for interactions which occur on another material.
The average purity for the water-in (water-out) configuration is $48.7 \%$ ($46.1 \%$) for all events with a two photon invariant mass less than 500~\MeVcc{} corresponding to a rejection of more than $99.5 \%$ of the background events. 
Figure~\ref{fig:effmom} shows the efficiency of the \ncpi{} selection as a function of the true \pizero{} momentum.

\section{\label{sec:selectedsample} Selected Event Samples}

\begin{table}
  \centering
  \caption{\label{tab:finalsumwi} The breakdown of the sample of events that satisfy the selection criteria for the P\O{}D water-in configuration.  
The sample of simulated events is broken down into signal and background and then the background sample is further subdivided by interaction type.
The values have been rounded.}
  \begin{ruledtabular}
    \begin{tabular}{ lcccccc} 
      & \multicolumn{3}{c}{Signal-Enriched Sample} & \multicolumn{3}{c}{Background-Enriched Sample} \\
      \hline
      Data & \DataWaterInSelected{} & & & \DataWaterInMuDk{} & & \\ 
      Expectation & \MCWaterInSelected{} & & & \MCWaterInMuDk{} & & \\ 
      \hline 
      \tableIndentOne{}Signal & & $435$ & & & $33$ & \\
      \tableIndentOne{}Background & & $458$ & & & $297$ & \\
      \hline
      \tableIndentTwo{}Neutral Current & & & $109$ & & & $74$ \\
      \tableIndentTwo{}Charged Current w/ \pizero{} & & & $56$ & & & $39$ \\
      \tableIndentTwo{}Charged Current w/o \pizero{} & & & $239$& & & $167$ \\ 
      \tableIndentTwo{}External & & & $39$ & & & $9$ \\
      \tableIndentTwo{}Multiple & & & $15$ & & & $8$ \\ 
    \end{tabular}
  \end{ruledtabular}
\end{table}

\begin{table}
  \centering
  \caption{\label{tab:finalsumwo} The breakdown of the sample of events that satisfy the selection criteria for the P\O{}D water-out configuration.  
The sample of simulated events is broken down into signal and background and then the background sample is further subdivided by interaction type.
The values have been rounded.}
  \begin{ruledtabular}
    \begin{tabular}{ lcccccc } 
      & \multicolumn{3}{c}{Signal-Enriched Sample} & \multicolumn{3}{c}{Background-Enriched Sample} \\
      \hline
      Data & \DataWaterOutSelected{} & & 
           & \DataWaterOutMuDk{} & & \\ 
      Expectation & \MCWaterOutSelected{} & &  
                              & \MCWaterOutMuDk{} & & \\
      \hline 
      \tableIndentOne{}Signal & & $290$ & & & $24$ & \\ 
      \tableIndentOne{}Background & & $339$ & & & $187$ & \\
      \hline
      \tableIndentTwo{}Non-Signal Neutral Current & & & $68$ & & & $45$ \\ 
      \tableIndentTwo{}Charged Current w/ \pizero{} & & & $40$ & & & $22$ \\ 
      \tableIndentTwo{}Charged Current w/o \pizero{} & & & $150$ & & & $100$ \\
      \tableIndentTwo{}External & & & $70$ & & & $13$ \\ 
      \tableIndentTwo{}Multiple & & & $11$ & & & $7$ \\ 
    \end{tabular}
  \end{ruledtabular}
\end{table}

\begin{figure}
  \centering
  \subfigure[Water-In configuration]{      \includegraphics[width=0.45\textwidth]{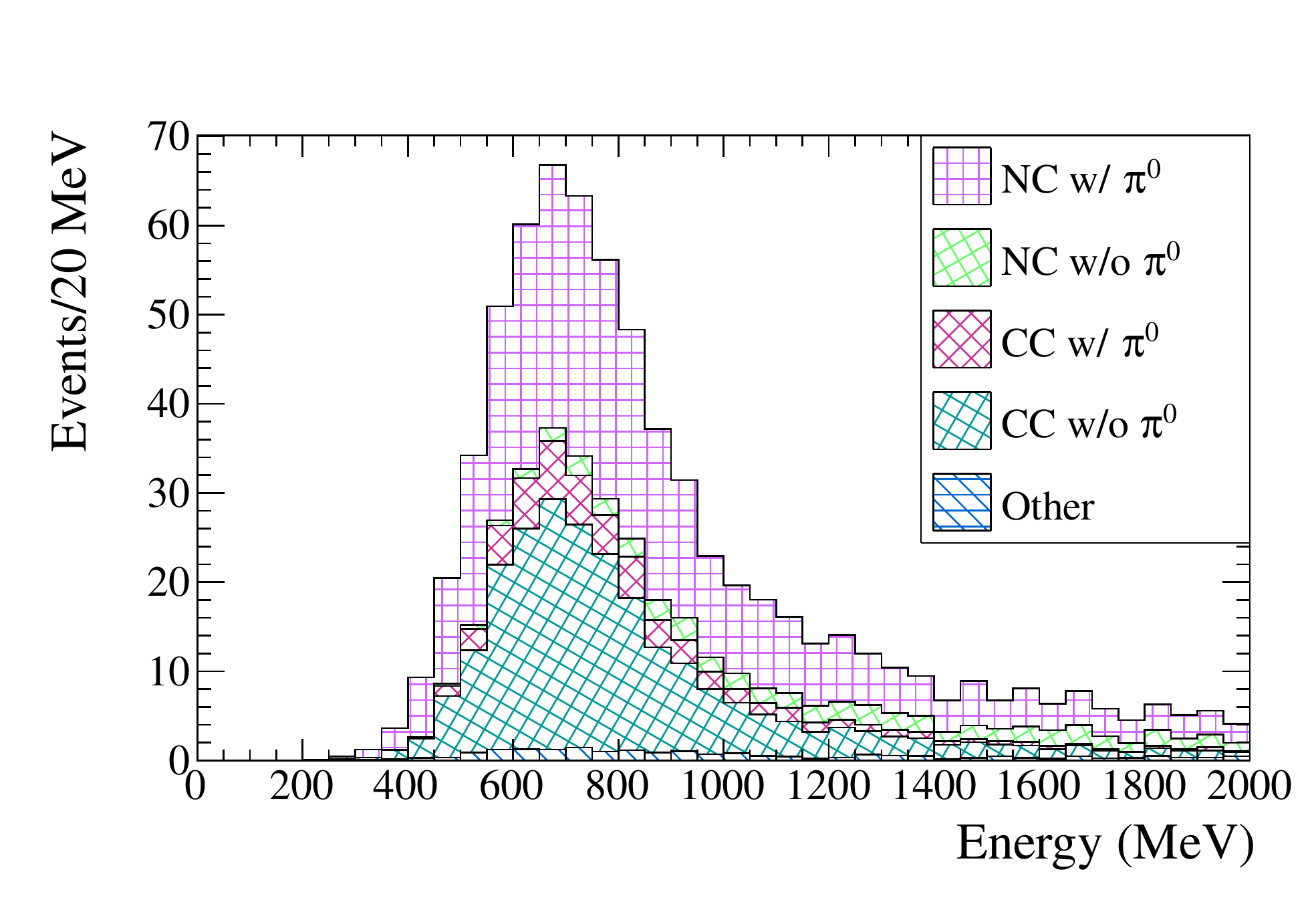}}
  \subfigure[Water-Out configuration]{ \includegraphics[width=0.45\textwidth]{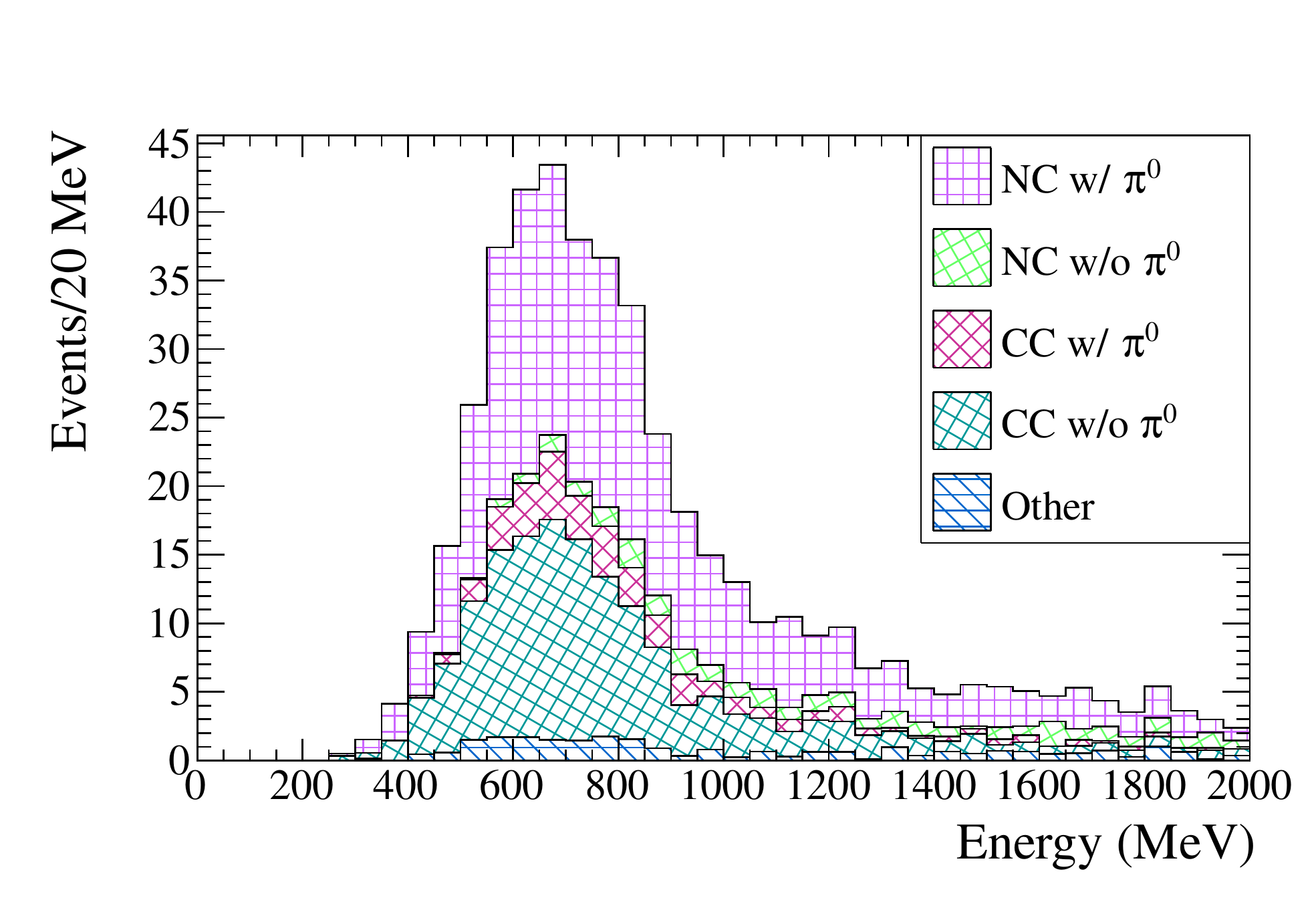}}
  \caption{\label{fig:neuten} The distribution of the true neutrino
    energy for the simulated events in the signal-enriched sample.}
\end{figure}

Tables~\ref{tab:finalsumwi} and~\ref{tab:finalsumwo} show the number of observed and expected events found in the signal-enriched and background-enriched samples.
The expectation for each sample is broken down into the number of expected signal and background events, and the number of background events is further broken down by the presence of charged leptons with and without a \pizero{} in the final state of the neutrino interaction.
Categories are also included for simulated events containing multiple neutrino interactions, and background entering from outside the P\O{}D.
Approximately $10 \%$ of the events in the background-enriched sample are due to signal interactions.
In the data, \DataWaterInSelected{} events were selected as an \ncpi{}-enriched sample for the water-in configuration and \DataWaterOutSelected{} events were selected for the water-out configuration of the P\O{}D compared to an expectation of \MCWaterInSelected{} (\MCWaterOutSelected{}) for the water-in (water-out) configuration.  
The distribution of the true neutrino energy for the selected sample of simulated events is shown in Figure~\ref{fig:neuten} separated by event topology with the mean neutrino energy for the \ncpi{} signal being \NCPiNeutrinoEnergy{}.

\begin{figure}
  \centering
  \subfigure[Water-In configuration]{ \includegraphics[width=0.45\textwidth]{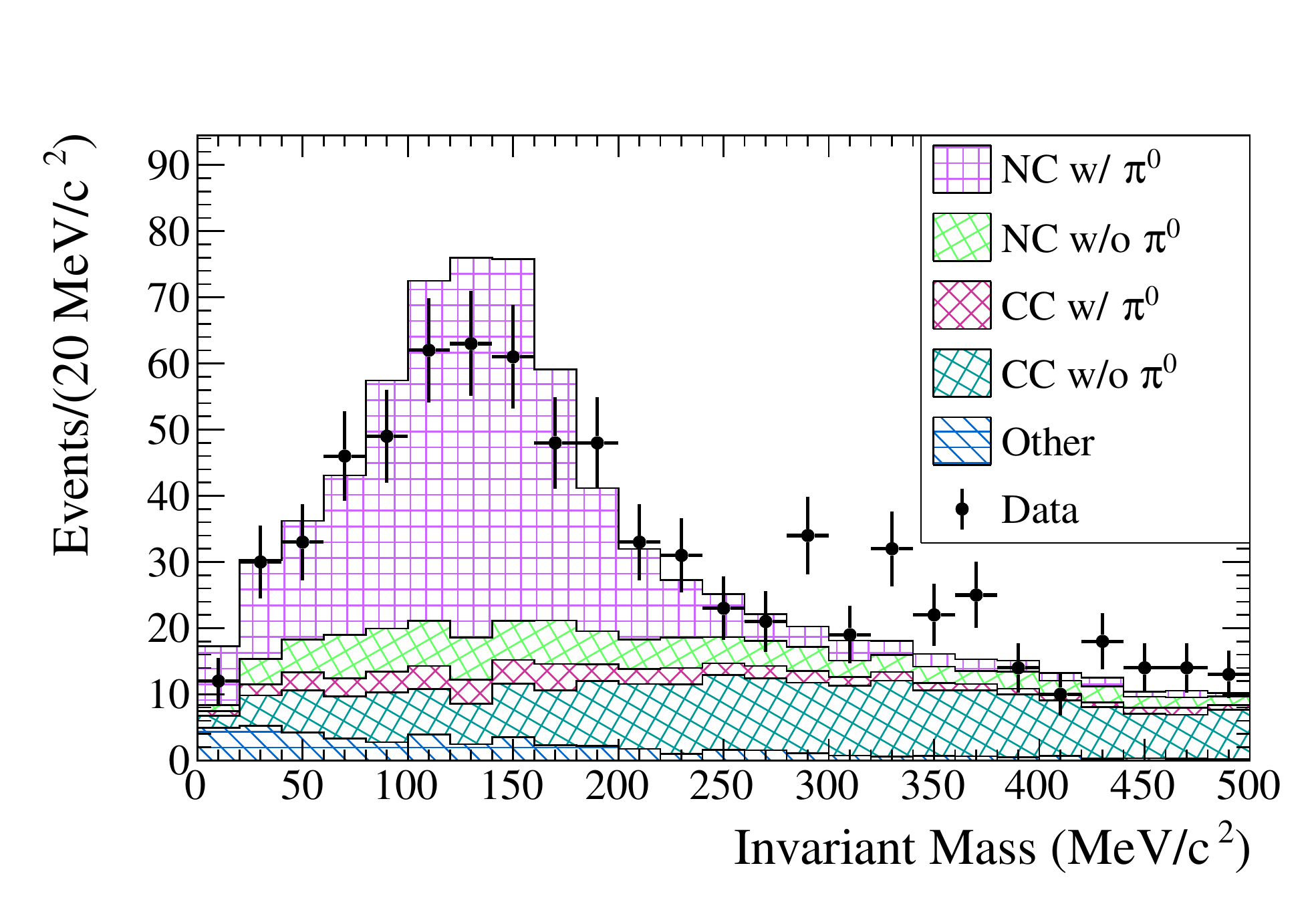}
\includegraphics[width=0.45\textwidth]{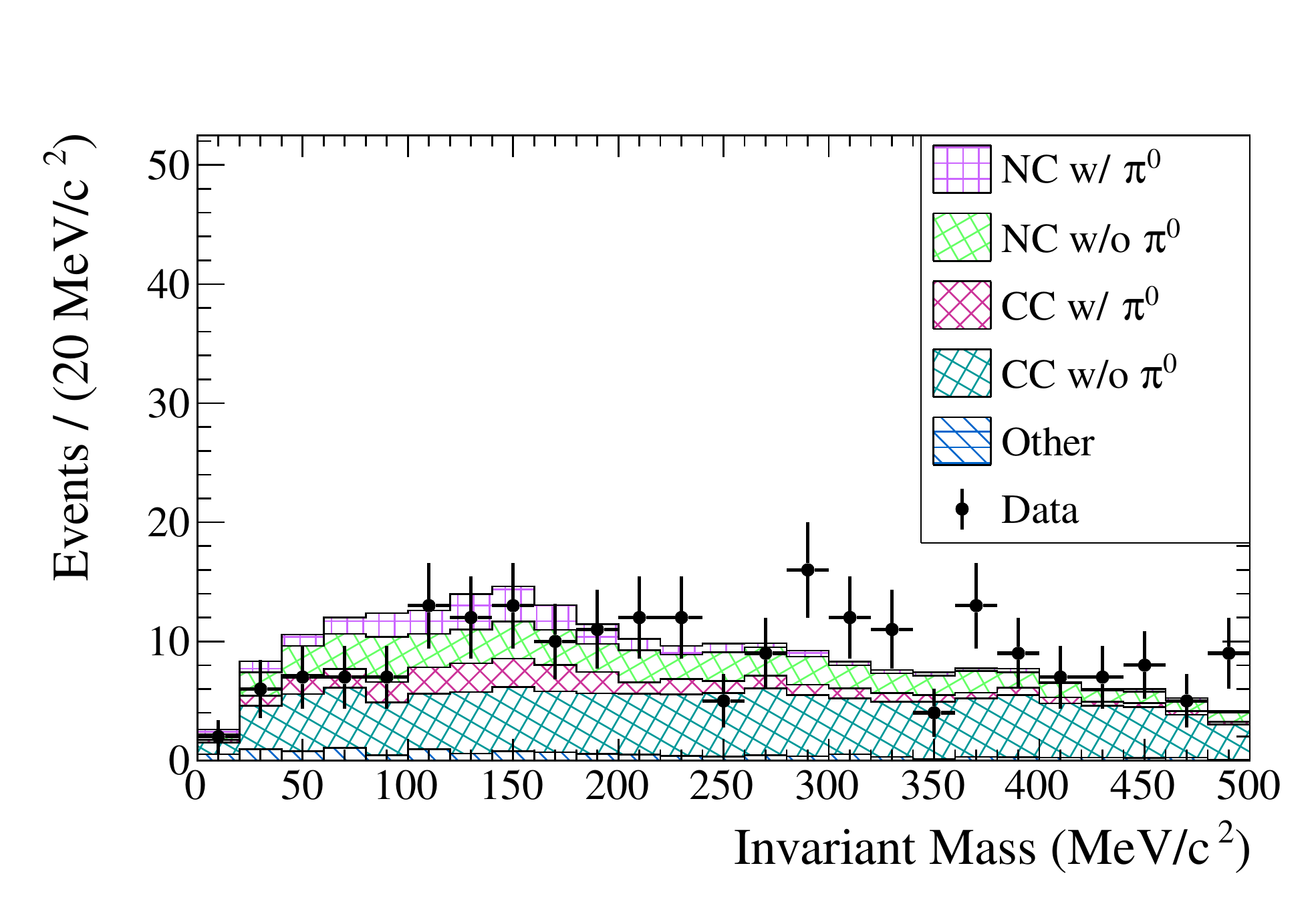}}
  \subfigure[Water-Out configuration]{ \includegraphics[width=0.45\textwidth]{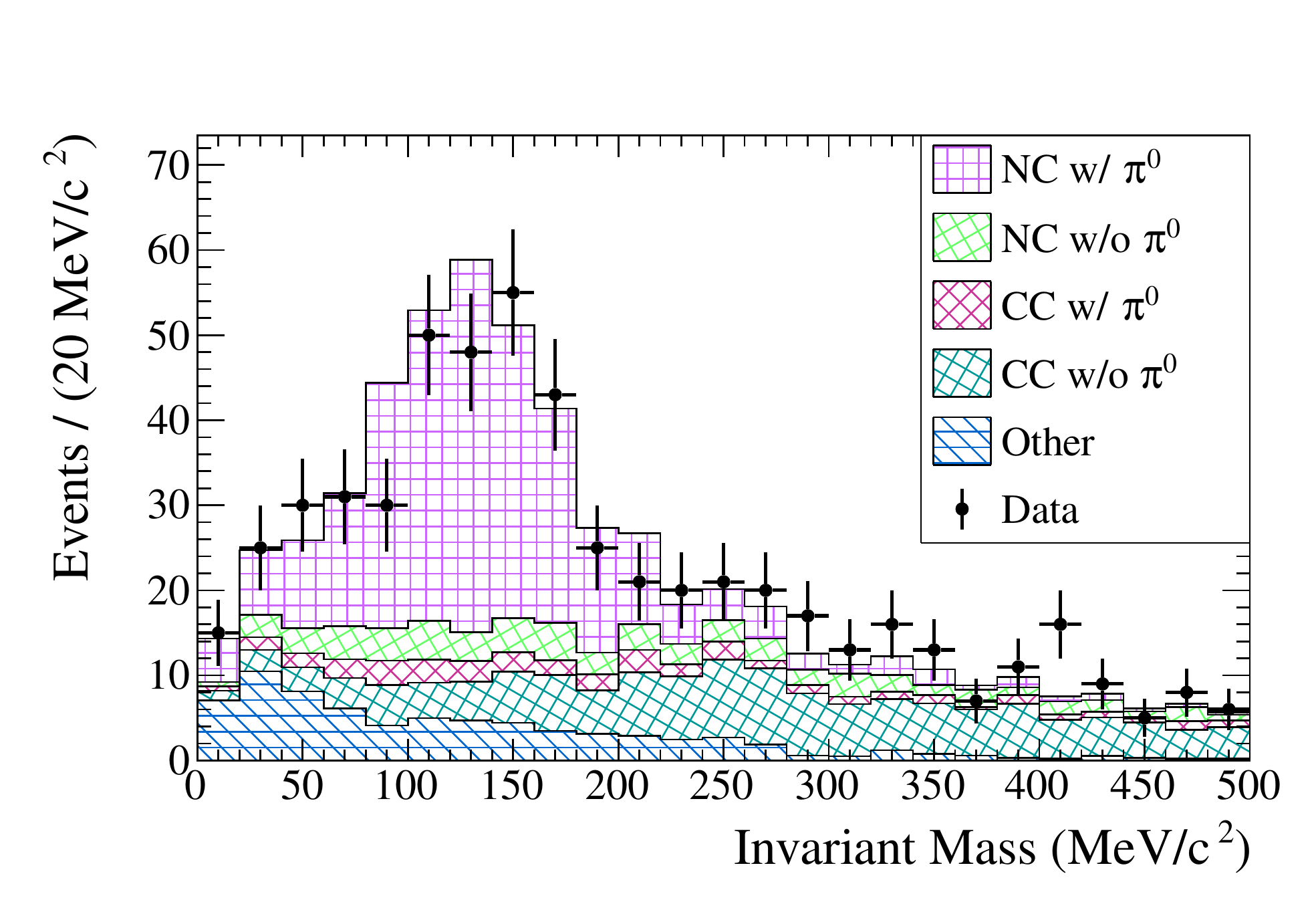}
\includegraphics[width=0.45\textwidth]{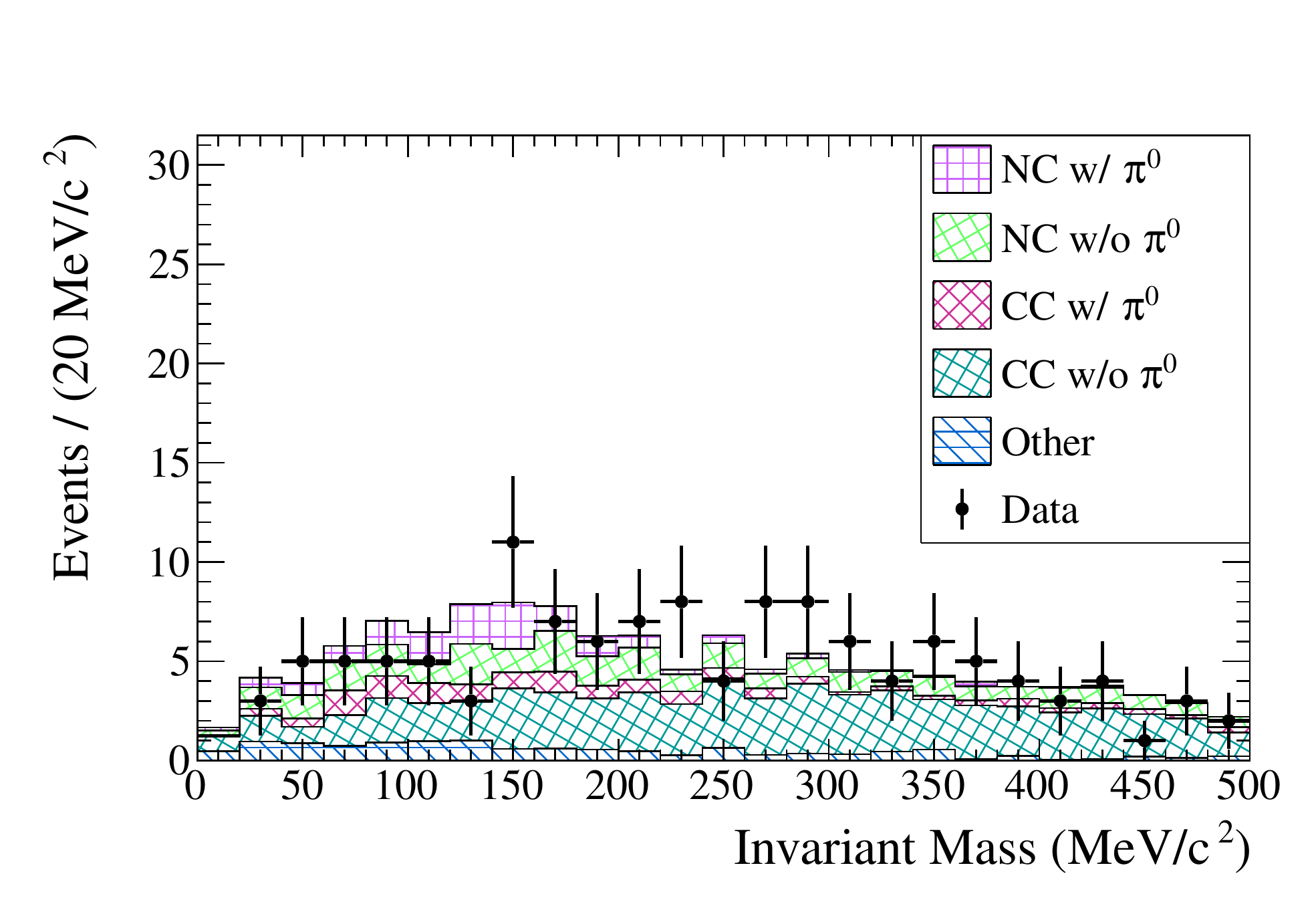}}
  \caption{\label{fig:invmass} The expected and observed reconstructed invariant mass for the signal-enriched and background-enriched samples in both the water-in and water-out configurations.  The left (right) plots show the distributions for the signal-enriched (background-enriched) samples.  The expectation for each sample is normalized to the observed number of events.}
\end{figure}

\begin{figure}
  \centering
  \subfigure[Water-In configuration]{ \includegraphics[width=0.45\textwidth]{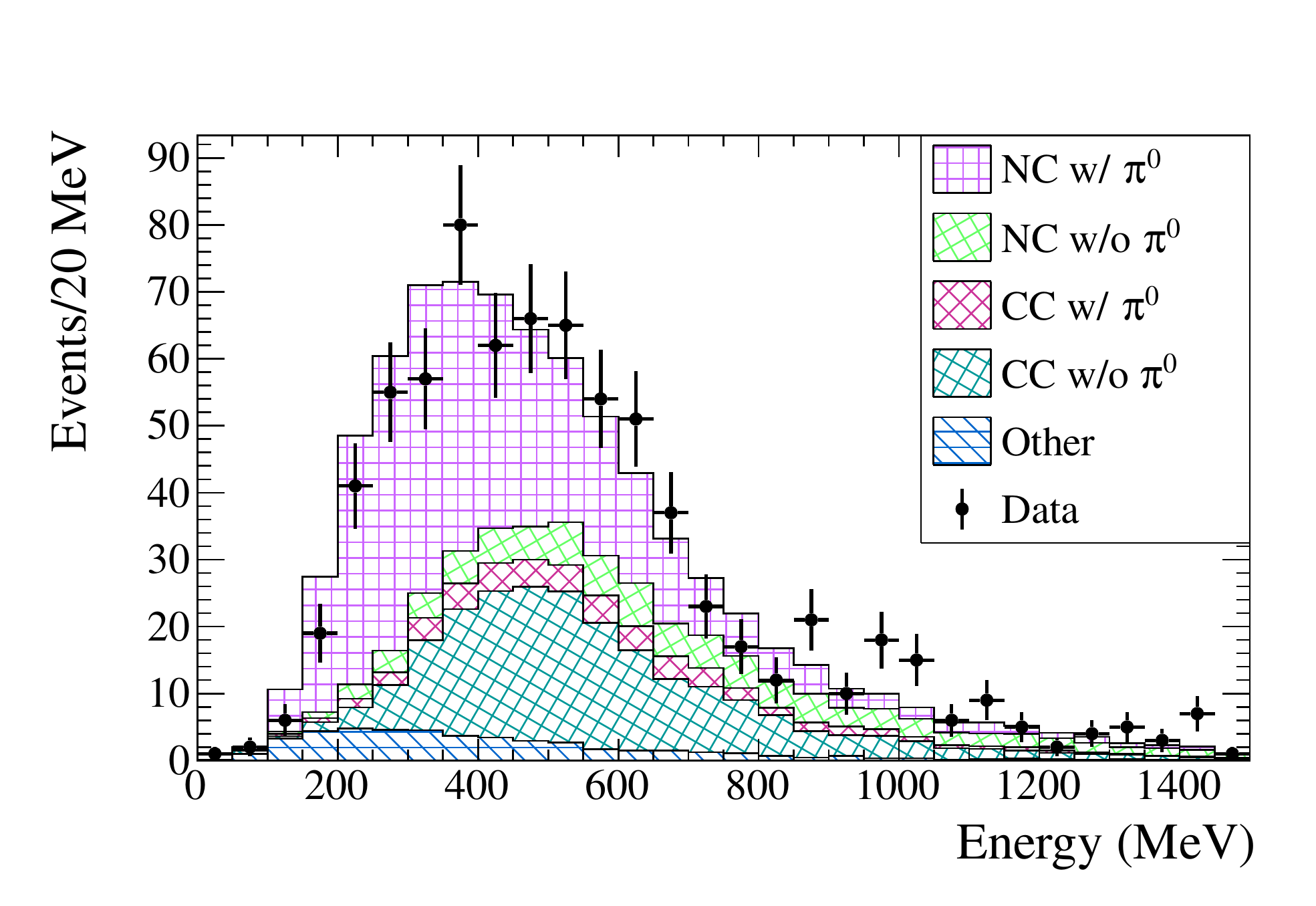}}
  \subfigure[Water-Out configuration]{
    \includegraphics[width=0.45\textwidth]{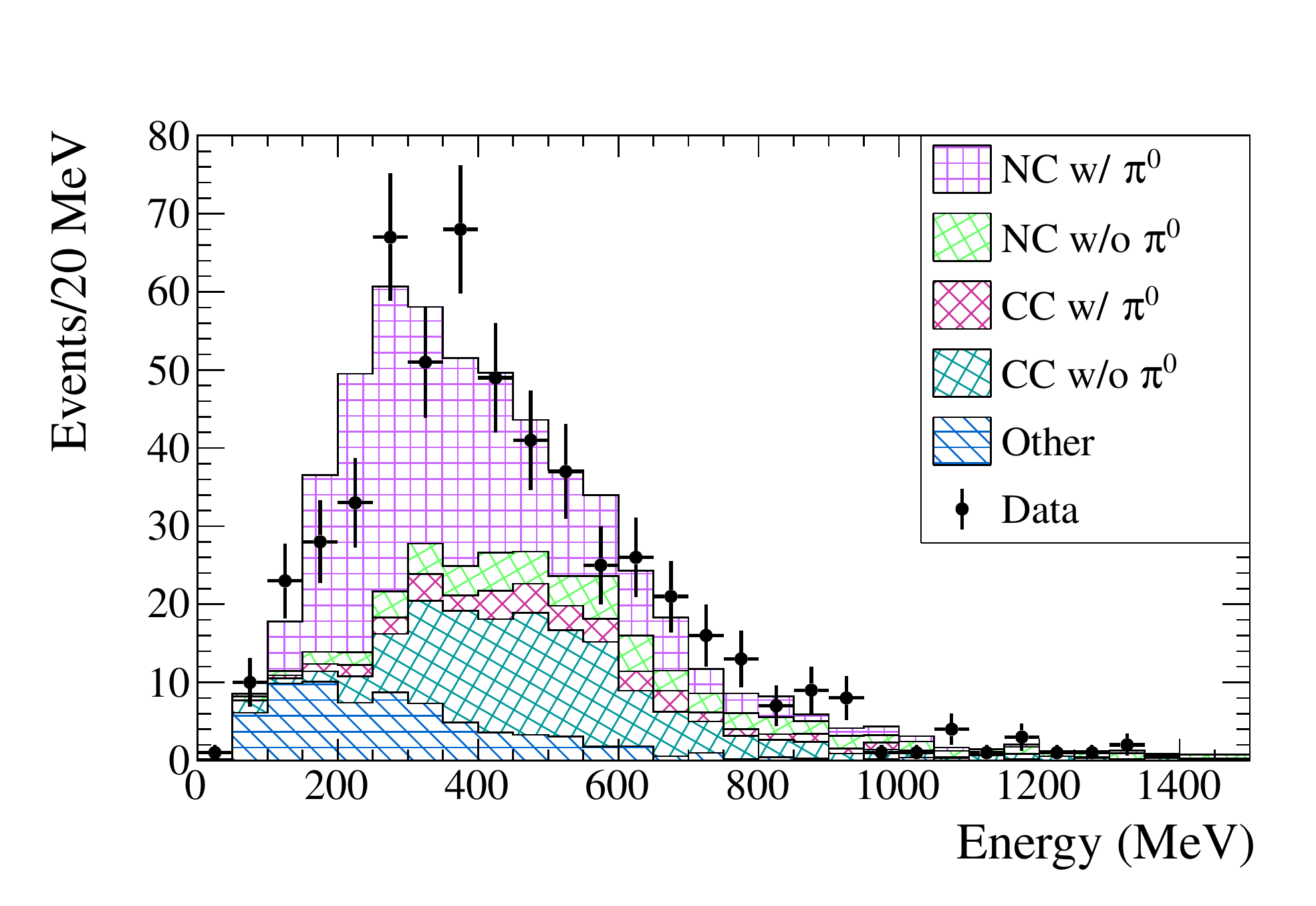}
  }
  \caption{\label{fig:recopi0energy} The expected and observed reconstructed \pizero{} energy for the signal-enriched sample.  The expectation for each sample is normalized to the observed number of events.}
\end{figure}

The event signature for this analysis is two reconstructed photons with an invariant mass, $M$, close to that of the \pizero{}.
The reconstructed invariant mass is  
$M = \sqrt{2E_{1}E_{2}(1-\cos{\theta})}$
where $E_{1}$ and $E_2$ are the reconstructed energies of each photon candidate and $\theta$ is the angle between the photon candidates.
Figure~\ref{fig:invmass} shows the distribution of the selected events in the signal-enriched and background-enriched samples where the expectation for each distribution has been normalized to the observed number of events.
The reconstructed energy distribution of the signal-enriched samples is shown in Figure~\ref{fig:recopi0energy}.
The expected composition for each distribution is shown using the same breakdown as in Tables~\ref{tab:finalsumwi} and~\ref{tab:finalsumwo}, however, the contributions from external and multiple interactions have been combined into a single category.

\section{\label{sec:signalextraction} Extracting the Signal Event Rate}
The number of \ncpi{} events is found using a six parameter unbinned extended maximum likelihood fit to the invariant mass distribution of the signal-enriched and background-enriched samples.  
Four of the parameters, $N_\textrm{Sig}^\textrm{SE}$, $N_\textrm{Bkg}^\textrm{SE}$, $N_\textrm{Sig}^\textrm{BE}$, and $N_\textrm{Bkg}^\textrm{BE}$, are related to the number of signal and background events in the signal-enriched (SE) and background-enriched (BE) samples.
The remaining two parameters control the energy scale of electromagnetic particles relative to minimum ionizing tracks, and the shape of the expected background.

The likelihood is extended by assuming that the probability of the observed numbers of signal-enriched ($N_{\gamma\gamma}^{\textrm{SE}}$) and background-enriched ($N_{\gamma\gamma}^{\textrm{BE}}$) events is given by the product of Poisson distributions and, in each case, the expected number of two photon events is a sum of the signal events ($N_\textrm{Sig}$) and the background events ($N_\textrm{Bkg}$).
The expected invariant mass distribution for the signal (background) events in the signal-enriched (background-enriched) sample is generated using the simulation after event reconstruction.  
The distributions are normalized such that the sum over the signal-enriched bins is equal to the total number of events, $N_{\gamma\gamma}^{\textrm{SE}}$, and, likewise, for the background-enriched sample, $N_{\gamma\gamma}^{\textrm{BE}}$.

\begin{table}
  \centering
  \caption{\label{tab:FitResultsFixG} 
Best fit values for the number of events in the P\O{}D water-in and P\O{}D water-out configurations.
Only statistical uncertainties are included.}
  \begin{ruledtabular}
    \begin{tabular}{l c c c c}
      & $N_\textrm{Sig}^\textrm{SE}$ (expected) 
            & $N_\textrm{Bkg}^\textrm{SE}$ (expected) 
            & $N_\textrm{Sig}^\textrm{BE}$ (expected) 
            & $N_\textrm{Bkg}^\textrm{BE}$ (expected)\\
      \hline
      Water-In & \DataWaterInSigFit{} (\MCWaterInSig{}) 
              & $338 \pm{} 26$ ($429$) 
              & $26.9 \pm{} 2.6$ ($33$) 
              & $245 \pm{} 15$ ($278$) \\
      Water-Out & \DataWaterOutSigFit{} (\MCWaterOutSig{}) 
            & $271 \pm{} 22$ ($335$) 
            & $20.4 \pm{} 2.2$ ($24$) 
            & $141 \pm{} 11$ ($184$) \\
    \end{tabular}
  \end{ruledtabular}
\end{table}

\begin{figure}
  \centering
  \subfigure[Water-In configuration]{ \label{fig:FitResults-in}\includegraphics[width=0.90\textwidth]{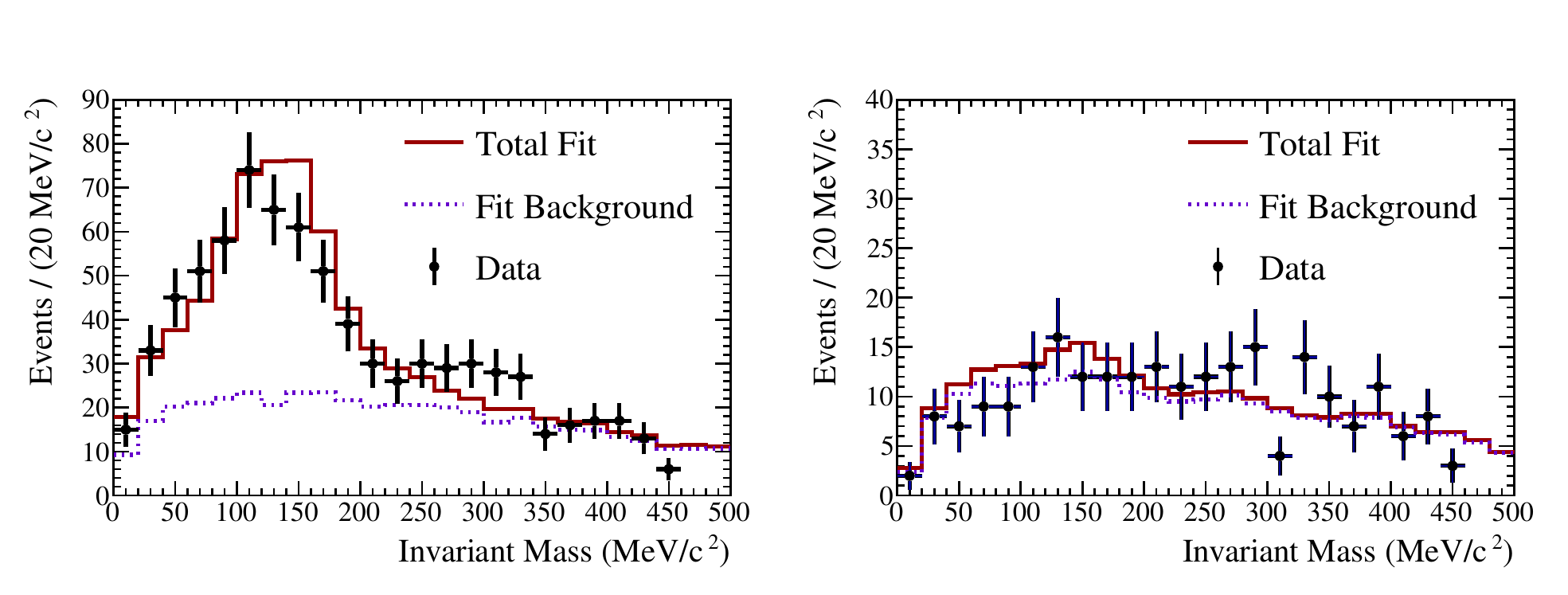}}
  \subfigure[Water-Out configuration]{  \label{fig:FitResults-out}\includegraphics[width=0.90\textwidth]{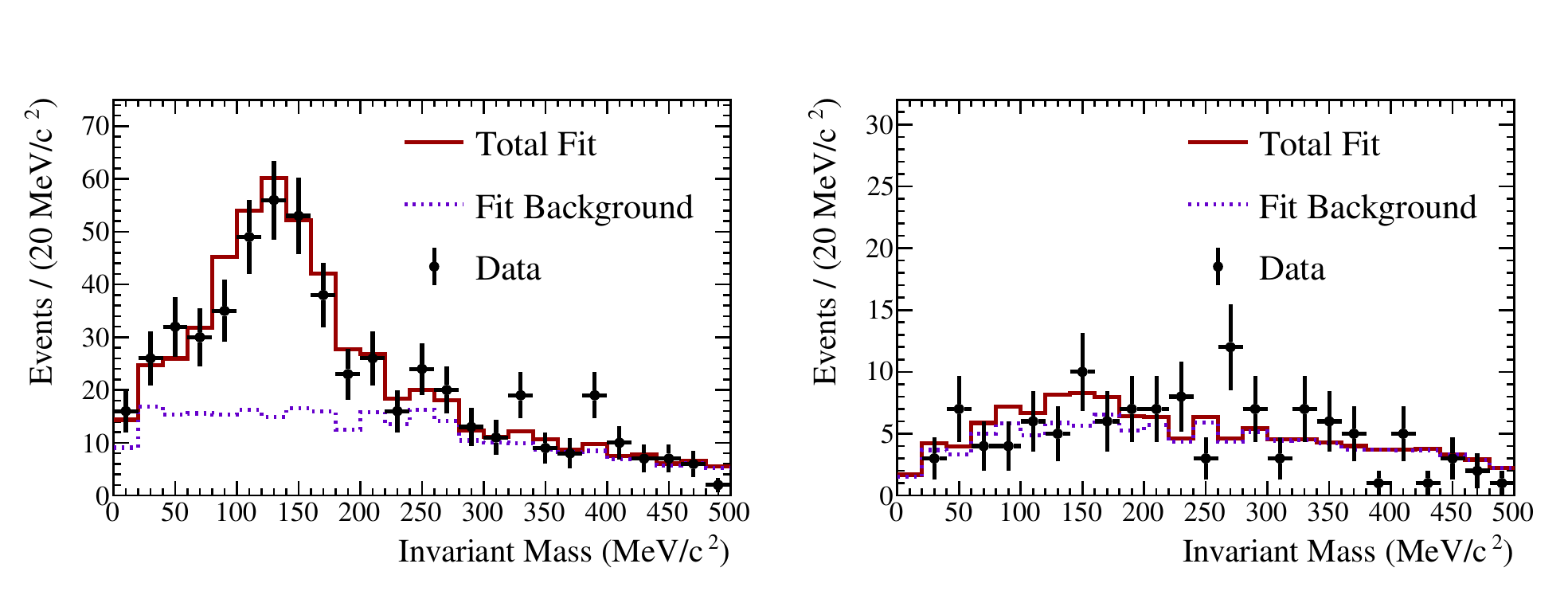}}
  \caption{\label{fig:FitResults} Comparison of the data to the best fit invariant mass distributions for the water-in and water-out configurations, with the best fit energy scale applied to the data.  The left (right) plots show the distributions for the signal-enriched (background-enriched) samples. The effect of the measured energy scale on the 500~\MeVcc{} selection criterion has not been applied to the expectation.
}
\end{figure}

The ratio of the number of signal and background events in the signal-enriched sample to the numbers in the background-enriched sample ($N_\textrm{Sig}^\textrm{SE} / N_\textrm{Sig}^\textrm{BE}$ and $N_\textrm{Bkg}^\textrm{SE} / N_\textrm{Bkg}^\textrm{BE}$) 
are determined by the efficiency of the muon decay tag and the probability of a muon decay tag false positive.
Both relations are estimated using a sample of stopping muons from neutrino interactions occurring upstream of the P\O{}D.
This sample is selected by requiring a single track-like object entering the upstream face of the detector, and stopping in the water target region.

The number of background events in the signal-enriched sample is related to the background events in the background-enriched sample by the muon decay reconstruction efficiency and is allowed to vary within the uncertainty on the muon decay tag efficiency.
For the water-in (water-out) configuration, the expected efficiency is $45.6 \%$ ($43.9 \%$) and the observed efficiency is $44.1\pm{}0.5 \%$ ($46.2\pm{}0.6 \%$).
The fractional difference between data and expectation is combined with its statistical error and used as a Gaussian constraint in the likelihood on the ratio between the number of background events in the signal-enriched and background-enriched samples.
The constraint is \WaterBpBConstraint{} for the water-in configuration and \AirBpBConstraint{} for the water-out configuration.

The fitted number of signal events in the background-enriched sample relative to the number in the signal-enriched sample is allowed to vary within the uncertainty on the probability of a false positive muon decay tag.
The uncertainty in modeling the false muon decay tag rate has been estimated by fitting the time distribution of muon decay tags occurring after a stopping muon, but within the same trigger window, to an exponential plus a constant.
The fitted exponential lifetimes are consistent with the expectation for muon decay, and the constant term estimates the probability of incorrectly finding a muon decay tag.
For the water-in (water-out) configuration there is a $1.1 \pm{} 0.5 \%$ ($1.3 \pm{} 0.7 \%$) difference in the constant term between the data and expectation which provides a \WaterSpSConstraint{} (\AirSpSConstraint{}) constraint on the false tag probability.

Since there is an uncertainty in the shape of the background underneath the \pizero{} invariant mass peak, an extra shape parameter has been added to the fit.
The deviation from the expected background shape is assumed to have the same shape as the signal probability distribution, while the normalization is constrained by the background-enriched sample.
The shape factor is allowed to be positive or negative meaning that the amount of background in the region of the \pizero{} invariant mass can be either increased or decreased.
Two cases are considered in the fit.
In the first instance, the number of signal and background events are determined by using the nominal shape for the background which is equivalent to fixing the shape parameter to a value of zero.
In the second case, no prior constraint is placed on the shape parameter, and the uncertainty in the rate is estimated by constraining it with both the signal-enhanced and background-enhanced samples.  See Section~\ref{sec:systematics} where this case is used to estimate the systematic uncertainty due to the unknown background shape.

The overall energy scale in the P\O{}D is set using penetrating muon tracks, and must be translated to an energy scale for electromagnetic particles with uncertainty introduced due to the relative response of the detector to different particle types.
The final electromagnetic energy scale is determined using the position of \pizero{} invariant mass peak.
The mean difference between the reconstructed and true photon opening angle in the simulation is small ($0.01$~rad) and has negligible effect on the invariant mass distribution so the difference between the measured and simulated invariant mass scales is assigned to the energy scale uncertainty.
No prior constraint is placed on the energy scale parameter, however, based on a survey of the detector material distribution and the uncertainties in the particle propagation model, the prior uncertainty is approximately $10 \%$ relative to the energy scale determined using penetrating muons.
In the water-in (water-out) configuration, the fitted value for the electromagnetic energy scale parameter is $89.5\pm{}3.4 \%$ ($96.7\pm{}0.6 \%$).

The best fit values for the number of signal and background events with the energy scale parameter unconstrained, while using the nominal shape for the background, are shown in Table~\ref{tab:FitResultsFixG}.  
Figure~\ref{fig:FitResults} compares the invariant mass expectation to the data, where the energy scale correction determined during the fit has been applied to the data.
Because the criteria requiring events have a reconstructed invariant mass less than $500~\MeVcc{}$ is applied prior to determining the best fit energy scale, the mass bins above  $440~\MeVcc{}$ are not fully populated with data.
The goodness of fit is calculated as a binned $\chi^2$ of the invariant mass distributions between $0$--$440~\MeVcc{}$ where the range has been limited to the region that the data populates.
Considering only statistical uncertainty, the $\chi^2$ value for the P\O{}D water-in configuration is $40.4$ for 39 degrees of freedom, leading to a p-value of $0.41$.
The $\chi^2$ value for the P\O{}D water-out configuration is $53.5$ for 39 degrees of freedom, leading to a p-value of $0.06$.

\section{Systematics}\label{sec:systematics}
The systematic uncertainties are summarized in Table~\ref{tab:sysErrors}, and are described below.
The detector systematics are separately estimated for the water-in and water-out configurations.  
Since the detector performance is different, and run periods do not overlap in time, the systematic uncertainty related to detector performance is assumed to be uncorrelated between the water-in and water-out configurations.

\begin{table}
  \centering
  \caption{\label{tab:sysErrors} 
Summary of event rate systematic errors.
The uncorrelated systematic errors are tabulated and summed, followed by the systematic error that is correlated between the water-in and water-out configurations.}
  \begin{ruledtabular}
    \begin{tabular}{lcc}
      Parameter & \multicolumn{2}{c}{Uncertainty}\\
      & Water-In & Water-Out \\
      \hline
      Geometry Differences & \GeometryVarSys{} & \GeometryVarSys{} \\
      PE Peak Discrepancy & \WaterInPEPeakSys{} & \WaterOutPEPeakSys{} \\
      Energy Scale & \WaterInEnergyScaleSys{} & \WaterOutEnergyScaleSys{} \\
      Channel to Channel Variations & $< 0.1 \%$ & $<0.1 \%$ \\
      Time Variation of Energy Scale & \TimeVariationSys{} & \TimeVariationSys{} \\
      Mass Uncertainty & \WaterInMassSys{} & \WaterOutMassSys{} \\
      Alignment & \AlignmentSys{} & \AlignmentSys{} \\
      Fiducial Volume Scaling & \WaterInFiducialVolumeSys{} & \WaterOutFiducialVolumeSys{} \\
      Fiducial Volume Shift & \WaterInFiducialShiftSys{} & \WaterOutFiducialShiftSys{} \\
      Flux Shape and Event Generator & \WaterInFluxSys{} & \WaterOutFluxSys{} \\
      Track PID Efficiency & \WaterInTrackPIDSys{} & \WaterOutTrackPIDSys{} \\
      Shower PID Efficiency & \WaterInShowerPIDSys{} & \WaterOutShowerPIDSys{} \\
      Object Separation & \WaterInShowerSepSys{} & \WaterOutShowerSepSys{} \\
      Charge-In-Shower & \WaterInShowerChargeSys{} & \WaterOutShowerChargeSys{} \\
      Background Shape (statistical) & \WaterInGFactorStat{} & \WaterOutGFactorStat{} \\
      \hline
      Total Uncorrelated Systematic & \WaterInTotalUncorrelatedSys{} & \WaterOutTotalUncorrelatedSys{} \\
      \hline
      Total Correlated Systematic & \multicolumn{2}{c}{\TotalCorrelatedSys{}} \\
      \hline
      Total Systematic & \WaterInTotalSystematic{} & \WaterOutTotalSystematic{} \\
    \end{tabular}
  \end{ruledtabular}
\end{table}

Because the event reconstruction proceeds in two stages, first reconstructing track-like signatures in each event, and then reconstructing the remaining activity assuming showering signatures, reconstruction efficiencies primarily affect the result in two ways.
First, an inefficiency is introduced when an electromagnetic object is reconstructed as track-like, because the object will then not be considered by the shower reconstruction.
Other efficiencies are more closely related to the shower reconstruction, including efficiencies related to the particle identification of showering signatures, the reconstructed distance between showering objects, and the fraction of the visible energy assigned to each showering object.

The track particle identification efficiency uncertainty is estimated using the sample of stopping muons described in Section~\ref{sec:signalextraction}.
The uncertainty for each input parameter to the particle identification procedure is estimated and propagated through the particle identification likelihood to determine the effect on the identification efficiency.
For simulated muons, there is a $5.40\pm{}0.05 \%$ ($5.06\pm{}0.03 \%$) uncertainty in the misidentification rates for the water-in (water-out) configuration.
Combining the difference in quadrature with the statistical uncertainty leads to a total track object particle identification systematic of \WaterInTrackPIDSys{} (\WaterOutTrackPIDSys{}) for the water-in (water-out) configuration.

The efficiencies of the charge-in-shower, object separation and shower particle identification criteria are related to the properties of a showering particle and are studied using control samples selected by reversing these cuts to create double ``side-band'' distributions.
For example, to estimate the uncertainty in the efficiency of the charge-in-shower criterion, events that fail the object separation and shower particle identification criteria, but which pass all other criteria, are selected to create a control sample with low signal purity.
The estimated uncertainty for the efficiency of the charge-in-shower criterion is then the relative difference between the percentage of control sample events passing the criterion relative to the expectation.
For the water-in configuration, $54.0 \%$ of the simulated, and $51.3\pm{}2.2 \%$ of the data control sample events are selected, combining the difference and the statistical error in quadrature leads to a systematic uncertainty of \WaterInShowerChargeSys{} in the efficiency due to the charge-in-shower criterion.
A similar calculation is done for the water-out configuration.
The procedure is then repeated for the object separation and shower particle identification criteria.
Because these three uncertainties are estimated using statistically limited data sets collected during independent water-in and water-out run periods they are assumed to be uncorrelated between the configurations, and the uncertainty will likely be reduced by the collection of additional data.
The on-water uncertainty for these uncertainties is estimated by combining the water-in and water-out uncertainties in quadrature (summarized in Table~\ref{tab:sysErrors}).

After the best fit values were found in Section~\ref{sec:signalextraction}, the fitted value and uncertainty on the energy scale were used to estimate the effect of the energy scale on the estimated efficiency.
This effect was modeled by scaling the \ncpi{} reconstruction efficiency shown in Figure~\ref{fig:effmom} using many trials of the energy scale parameter distributed according to the best fit parameter and statistical uncertainty.
The shifted efficiency curve represents a new expectation for the trial energy scale parameter and is used to estimate the expected number of saved signal events in the simulation.
The fractional shift and RMS of the distribution of the expected number of signal events are then added in quadrature to estimate the uncertainty due to the energy scale for the estimated efficiency.
The time variation of the energy scale was tracked using through-going minimum ionizing particles, and it introduces an efficiency uncertainty of \TimeVariationSys{} for both the water-in and water-out configurations.

Several systematic uncertainties were associated with the fiducial volume.
Uncertainties on measurements of the detector mass were used to reweight the selected events to extract the uncertainty due to fiducial mass.
The effect of alignment between detector elements on the efficiency was studied and found to be negligible ($< 0.1 \%$).
Additionally, there are two fiducial volume uncertainties.
One reflects how the result is affected by changing the fiducial volume definition, while the other quantifies the uncertainty due to a systematic shift between the simulated and true detector volumes.
When combined, the fiducial volume uncertainty is $1.9 \%$ for water-in and $2.6 \%$ for water-out.

The systematic uncertainty due to the background shape is estimated by comparing the effect on the fitted signal rate with the shape parameter fixed to when it is unconstrained.
Following the procedure outlined in Section~\ref{sec:signalextraction}, the best values for the shape parameter with the water-in and water-out configurations are found to be statistically consistent with the $\chi^2$ value for the invariant mass distribution being $47.5$ for $38$ degrees of freedom for the water-in configuration and $38.7$ for $38$ degrees of freedom for the water-out configuration.
Because the backgrounds in both the water-in and water-out configuration arise from the same physical processes, it is assumed that this uncertainty is fully correlated between the water-in and water-out configurations and the uncertainty is directly applied to the on-water signal event rate.
The change in the water-in and water-out event rates between the case where the shape parameter is fixed to zero and where the shape parameter is free leads to a \CorrelatedGFactorSys{} uncertainty in the on-water rate due to the background shape parameter.

While this analysis uses background-enriched control samples to minimize the uncertainty due to the model of the cross sections, changes in the generated event kinematics have an effect.
This was studied using the neutrino flux shape and cross section uncertainties detailed in~\cite{Abe:2012av} and~\cite{Abe:2014usb}.
For the water-in configuration, the flux shape and cross section introduce a \WaterInFluxSys{} uncertainty, and for the water-out configuration, a \WaterOutFluxSys{} uncertainty.
Because rate of observed events is consistent with the expectation, the full \NCPiNeutrinoFluxSys{} flux uncertainty is included and given as a separate uncertainty.

\section{\label{sec:eventrate} Number of Events on Water}
The number of \ncpi{} events in the P\O{}D measured using both the water-in and water-out configuration (Table~\ref{tab:FitResultsFixG}) can be used to determine the number of events occurring directly on water.
The measured number of \ncpi{} events with water in the P\O{}D is found to be \DataWaterInSig{} during an exposure of \DataAirPOT{}~POT, where the systematic uncertainty includes effects that are correlated between the water-in and water-out configurations.
The ratio between the observed and expected rate is \DataMCWaterInRatio{}.
Similarly, with water out of the P\O{}D, the measured number is \DataWaterOutSig{} for an exposure of \DataWaterPOT{}~POT, and the ratio between the observed and expected rate is \DataMCWaterOutRatio{}.
To allow direct comparison to the expected \ncpi{} event rate, the quoted ratios include neither the \NCPiNeutrinoFluxSys{} flux normalization uncertainty nor the \ncpi{} cross section uncertainty

The total number of events on water is found using a statistical subtraction by relating the event rate during the water-in exposure to the event rate during the water-out exposure.
The total number of signal events in the water-in (WI) configuration can be divided into two parts, $N_\textrm{WI} = N_\textrm{On-Water} + N_\textrm{NW}$, where $N_\textrm{NW}$ is the number of signal events that occur on targets other than water (referred to as ``not-water'' events) in the water-in configuration.
The number of not-water (NW) events, which is proportional to the number of water-out (WO) events, can be subtracted from the total number of on-water events by
\begin{equation}\label{eq:OnWaterCalculation}
  N_{\textrm{On-Water}} = N_{\textrm{WI}} 
  - \frac{\epsilon_\textrm{NW} \textrm{POT}_\textrm{WI}}
         {\epsilon_\textrm{WO} \textrm{POT}_\textrm{WO}}
  N_{\textrm{WO}},
\end{equation}
where the efficiencies, $\epsilon_\textrm{NW}$ and $\epsilon_\textrm{WO}$, are given in Table~\ref{tab:effsummary}, and the POT is given in Table~\ref{tab:datapot}.
After the subtraction in Equation~\ref{eq:OnWaterCalculation}, \DataOnWaterSig{} events were found, where the uncertainties that are correlated between the water-in and water-out configurations have been taken into account.
The simulation predicts \MCOnWaterSig{} true \ncpi{} events on water.

The ratio of the number of measured to number of predicted on-water events, including the correlated and uncorrelated uncertainties described in Section~\ref{sec:systematics} is \DataMCOnWaterFluxRatio{}, where the  \ncpi{} cross section uncertainties are excluded.

\section{Conclusion}
An on-water \ncpi{} rate measurement has been performed by combining data from a \DataWaterPOT{}~POT neutrino beam exposure of the T2K ND280 P\O{}D using a water-in configuration with a \DataAirPOT{}~POT exposure using a water-out configuration.
This is the first use of the subtraction method to measure neutral current event rates with the T2K near detector.

The signal event rates are found using an extended maximum likelihood fit to the reconstructed invariant mass for each sample in a range of 0-500~\MeVcc{}.
As described in Section~\ref{sec:eventselection}, the phase space of the analysis has been limited to the region where the P\O{}D has acceptance.
The analysis finds \DataWaterInSig{} (\DataWaterOutSig{}) signal events in the P\O{}D water-in (water-out) data compared to an expectation of \MCWaterInSig{} (\MCWaterOutSig{}) events.
Excluding the \NCPiNeutrinoFluxSys{} normalization and \ncpi{} cross section uncertainties, the resulting observed to expected ratios are \DataMCWaterInRatio{} for water-in and \DataMCWaterOutRatio{} for water-out configurations.
Subtracting the water-in and water-out samples after correcting for the different POT and reconstruction efficiencies yields \DataOnWaterSig{} signal events on water compared to an expectation of \MCOnWaterSig{} events and an on-water \ncpi{} production rate of \DataMCOnWaterFluxRatio{} relative to the NEUT expectation.
As noted in Section~\ref{sec:systematics}, the largest systematic errors, for example, the uncertainty in the background shape, are determined using statistically limited data sets and additional exposure is expected to reduce these uncertainties.
The observed event rates are consistent with the expectation and indicate that the event rate from neutral current \pizero{} production is not underestimated.
This provides confidence that the neutral current \pizero{} background to electron neutrino appearance in T2K is not underestimated.

{\it Acknowledgements}\textemdash
We thank the J-PARC staff for superb accelerator performance. We thank the 
CERN NA61/SHINE Collaboration for providing valuable particle production data.
We acknowledge the support of MEXT, Japan; 
NSERC (Grant No. SAPPJ-2014-00031), NRC and CFI, Canada;
CEA and CNRS/IN2P3, France;
DFG, Germany; 
INFN, Italy;
National Science Centre (NCN) and Ministry of Science and Higher Education, Poland;
RSF, RFBR, and MES, Russia; 
MINECO and ERDF funds, Spain;
SNSF and SERI, Switzerland;
STFC, UK; and 
DOE, USA.
We also thank CERN for the UA1/NOMAD magnet, 
DESY for the HERA-B magnet mover system, 
NII for SINET4, 
the WestGrid and SciNet consortia in Compute Canada, 
and GridPP in the United Kingdom.
In addition, participation of individual researchers and institutions has been further 
supported by funds from ERC (FP7), H2020 Grant No. RISE-GA644294-JENNIFER, EU; 
JSPS, Japan; 
Royal Society, UK; 
the Alfred P. Sloan Foundation and the DOE Early Career program, USA.

\bibliographystyle{apsrev4-1}
\bibliography{P0DPi02015}

\end{document}

%% file: author_list.tex

\newcommand{\INSTEE}{\affiliation{University of Bern, Albert Einstein Center for Fundamental Physics, Laboratory for High Energy Physics (LHEP), Bern, Switzerland}}
\newcommand{\INSTFE}{\affiliation{Boston University, Department of Physics, Boston, Massachusetts, U.S.A.}}
\newcommand{\INSTD}{\affiliation{University of British Columbia, Department of Physics and Astronomy, Vancouver, British Columbia, Canada}}
\newcommand{\INSTGA}{\affiliation{University of California, Irvine, Department of Physics and Astronomy, Irvine, California, U.S.A.}}
\newcommand{\INSTI}{\affiliation{IRFU, CEA Saclay, Gif-sur-Yvette, France}}
\newcommand{\INSTGB}{\affiliation{University of Colorado at Boulder, Department of Physics, Boulder, Colorado, U.S.A.}}
\newcommand{\INSTFG}{\affiliation{Colorado State University, Department of Physics, Fort Collins, Colorado, U.S.A.}}
\newcommand{\INSTFH}{\affiliation{Duke University, Department of Physics, Durham, North Carolina, U.S.A.}}
\newcommand{\INSTBA}{\affiliation{Ecole Polytechnique, IN2P3-CNRS, Laboratoire Leprince-Ringuet, Palaiseau, France }}
\newcommand{\INSTEF}{\affiliation{ETH Zurich, Institute for Particle Physics, Zurich, Switzerland}}
\newcommand{\INSTEG}{\affiliation{University of Geneva, Section de Physique, DPNC, Geneva, Switzerland}}
\newcommand{\INSTDG}{\affiliation{H. Niewodniczanski Institute of Nuclear Physics PAN, Cracow, Poland}}
\newcommand{\INSTCB}{\affiliation{High Energy Accelerator Research Organization (KEK), Tsukuba, Ibaraki, Japan}}
\newcommand{\INSTED}{\affiliation{Institut de Fisica d'Altes Energies (IFAE), The Barcelona Institute of Science and Technology, Campus UAB, Bellaterra (Barcelona) Spain}}
\newcommand{\INSTEC}{\affiliation{IFIC (CSIC \& University of Valencia), Valencia, Spain}}
\newcommand{\INSTEI}{\affiliation{Imperial College London, Department of Physics, London, United Kingdom}}
\newcommand{\INSTGF}{\affiliation{INFN Sezione di Bari and Universit\`a e Politecnico di Bari, Dipartimento Interuniversitario di Fisica, Bari, Italy}}
\newcommand{\INSTBE}{\affiliation{INFN Sezione di Napoli and Universit\`a di Napoli, Dipartimento di Fisica, Napoli, Italy}}
\newcommand{\INSTBF}{\affiliation{INFN Sezione di Padova and Universit\`a di Padova, Dipartimento di Fisica, Padova, Italy}}
\newcommand{\INSTBD}{\affiliation{INFN Sezione di Roma and Universit\`a di Roma ``La Sapienza'', Roma, Italy}}
\newcommand{\INSTEB}{\affiliation{Institute for Nuclear Research of the Russian Academy of Sciences, Moscow, Russia}}
\newcommand{\INSTHA}{\affiliation{Kavli Institute for the Physics and Mathematics of the Universe (WPI), The University of Tokyo Institutes for Advanced Study, University of Tokyo, Kashiwa, Chiba, Japan}}
\newcommand{\INSTCC}{\affiliation{Kobe University, Kobe, Japan}}
\newcommand{\INSTCD}{\affiliation{Kyoto University, Department of Physics, Kyoto, Japan}}
\newcommand{\INSTEJ}{\affiliation{Lancaster University, Physics Department, Lancaster, United Kingdom}}
\newcommand{\INSTFC}{\affiliation{University of Liverpool, Department of Physics, Liverpool, United Kingdom}}
\newcommand{\INSTFI}{\affiliation{Louisiana State University, Department of Physics and Astronomy, Baton Rouge, Louisiana, U.S.A.}}
\newcommand{\INSTJ}{\affiliation{Universit\'e de Lyon, Universit\'e Claude Bernard Lyon 1, IPN Lyon (IN2P3), Villeurbanne, France}}
\newcommand{\INSTHB}{\affiliation{Michigan State University, Department of Physics and Astronomy,  East Lansing, Michigan, U.S.A.}}
\newcommand{\INSTCE}{\affiliation{Miyagi University of Education, Department of Physics, Sendai, Japan}}
\newcommand{\INSTDF}{\affiliation{National Centre for Nuclear Research, Warsaw, Poland}}
\newcommand{\INSTFJ}{\affiliation{State University of New York at Stony Brook, Department of Physics and Astronomy, Stony Brook, New York, U.S.A.}}
\newcommand{\INSTGJ}{\affiliation{Okayama University, Department of Physics, Okayama, Japan}}
\newcommand{\INSTCF}{\affiliation{Osaka City University, Department of Physics, Osaka, Japan}}
\newcommand{\INSTGG}{\affiliation{Oxford University, Department of Physics, Oxford, United Kingdom}}
\newcommand{\INSTBB}{\affiliation{UPMC, Universit\'e Paris Diderot, CNRS/IN2P3, Laboratoire de Physique Nucl\'eaire et de Hautes Energies (LPNHE), Paris, France}}
\newcommand{\INSTGC}{\affiliation{University of Pittsburgh, Department of Physics and Astronomy, Pittsburgh, Pennsylvania, U.S.A.}}
\newcommand{\INSTFA}{\affiliation{Queen Mary University of London, School of Physics and Astronomy, London, United Kingdom}}
\newcommand{\INSTE}{\affiliation{University of Regina, Department of Physics, Regina, Saskatchewan, Canada}}
\newcommand{\INSTGD}{\affiliation{University of Rochester, Department of Physics and Astronomy, Rochester, New York, U.S.A.}}
\newcommand{\INSTHC}{\affiliation{Royal Holloway University of London, Department of Physics, Egham, Surrey, United Kingdom}}
\newcommand{\INSTBC}{\affiliation{RWTH Aachen University, III. Physikalisches Institut, Aachen, Germany}}
\newcommand{\INSTFB}{\affiliation{University of Sheffield, Department of Physics and Astronomy, Sheffield, United Kingdom}}
\newcommand{\INSTDI}{\affiliation{University of Silesia, Institute of Physics, Katowice, Poland}}
\newcommand{\INSTEH}{\affiliation{STFC, Rutherford Appleton Laboratory, Harwell Oxford,  and  Daresbury Laboratory, Warrington, United Kingdom}}
\newcommand{\INSTCH}{\affiliation{University of Tokyo, Department of Physics, Tokyo, Japan}}
\newcommand{\INSTBJ}{\affiliation{University of Tokyo, Institute for Cosmic Ray Research, Kamioka Observatory, Kamioka, Japan}}
\newcommand{\INSTCG}{\affiliation{University of Tokyo, Institute for Cosmic Ray Research, Research Center for Cosmic Neutrinos, Kashiwa, Japan}}
\newcommand{\INSTGI}{\affiliation{Tokyo Metropolitan University, Department of Physics, Tokyo, Japan}}
\newcommand{\INSTF}{\affiliation{University of Toronto, Department of Physics, Toronto, Ontario, Canada}}
\newcommand{\INSTB}{\affiliation{TRIUMF, Vancouver, British Columbia, Canada}}
\newcommand{\INSTG}{\affiliation{University of Victoria, Department of Physics and Astronomy, Victoria, British Columbia, Canada}}
\newcommand{\INSTDJ}{\affiliation{University of Warsaw, Faculty of Physics, Warsaw, Poland}}
\newcommand{\INSTDH}{\affiliation{Warsaw University of Technology, Institute of Radioelectronics, Warsaw, Poland}}
\newcommand{\INSTFD}{\affiliation{University of Warwick, Department of Physics, Coventry, United Kingdom}}
\newcommand{\INSTGE}{\affiliation{University of Washington, Department of Physics, Seattle, Washington, U.S.A.}}
\newcommand{\INSTGH}{\affiliation{University of Winnipeg, Department of Physics, Winnipeg, Manitoba, Canada}}
\newcommand{\INSTEA}{\affiliation{Wroclaw University, Faculty of Physics and Astronomy, Wroclaw, Poland}}
\newcommand{\INSTHE}{\affiliation{Yokohama National University, Faculty of Engineering, Yokohama, Japan}}
\newcommand{\INSTH}{\affiliation{York University, Department of Physics and Astronomy, Toronto, Ontario, Canada}}

\INSTEE
\INSTFE
\INSTD
\INSTGA
\INSTI
\INSTGB
\INSTFG
\INSTFH
\INSTBA
\INSTEF
\INSTEG
\INSTDG
\INSTCB
\INSTED
\INSTEC
\INSTEI
\INSTGF
\INSTBE
\INSTBF
\INSTBD
\INSTEB
\INSTHA
\INSTCC
\INSTCD
\INSTEJ
\INSTFC
\INSTFI
\INSTJ
\INSTHB
\INSTCE
\INSTDF
\INSTFJ
\INSTGJ
\INSTCF
\INSTGG
\INSTBB
\INSTGC
\INSTFA
\INSTE
\INSTGD
\INSTHC
\INSTBC
\INSTFB
\INSTDI
\INSTEH
\INSTCH
\INSTBJ
\INSTCG
\INSTGI
\INSTF
\INSTB
\INSTG
\INSTDJ
\INSTDH
\INSTFD
\INSTGE
\INSTGH
\INSTEA
\INSTHE
\INSTH

\author{K.\,Abe}\INSTBJ
\author{J.\,Amey}\INSTEI
\author{C.\,Andreopoulos}\INSTEH\INSTFC
\author{M.\,Antonova}\INSTEB
\author{S.\,Aoki}\INSTCC
\author{A.\,Ariga}\INSTEE
\author{Y.\,Ashida}\INSTCD
\author{S.\,Assylbekov}\INSTFG
\author{D.\,Autiero}\INSTJ
\author{S.\,Ban}\INSTCD
% ...

\collaboration{The T2K Collaboration}\noaffiliation